\documentclass[journal,twoside,web]{IEEEtran}
\usepackage{graphicx} \graphicspath{ {Images/} }
\usepackage{amsmath}
\usepackage{cite}
\usepackage{booktabs}
\usepackage{microtype}
\usepackage{balance}
\usepackage[colorlinks]{hyperref}

\title{RAVIR: A Dataset and Methodology for the Semantic Segmentation and Quantitative Analysis of Retinal Arteries and Veins in Infrared Reflectance Imaging}

\author{Ali Hatamizadeh,
        Hamid Hosseini,
        Niraj Patel,
        Jinseo Choi,
        Cameron C.~Pole,
        Cory M.~Hoeferlin,\\
        Steven D.~Schwartz
        and Demetri Terzopoulos
\thanks{Ali Hatamizadeh and Hamid Hosseini are co-primary authors.} \thanks{Corresponding author: Ali Hatamizadeh (ahatamiz@cs.ucla.edu).}
\thanks{Ali Hatamizadeh is with the Computer Science Department, Samueli School of Engineering, University of California, Los Angeles, CA, USA.}
\thanks{Hamid Hosseini, Niraj Patel, Jinseo Choi, Cameron C.~Pole, Cory M.~Hoeferlin and Steven D.~Schwartz are with the Stein Eye Institute, Department of Ophthalmology, David Geffen School of Medicine, University of California, Los Angeles, CA, USA.}
\thanks{Demetri Terzopoulos is with the Computer Science Department, Samueli School of Engineering, University of California, Los Angeles, CA, USA, and is a co-founder of VoxelCloud, Inc., Los Angeles, CA, USA.}
\thanks{This work was supported in part by an unrestricted grant from Research to Prevent Blindness (RPB) to the UCLA Department of Ophthalmology.}}

\begin{document}
\maketitle

\begin{abstract}
The retinal vasculature provides important clues in the diagnosis and monitoring of systemic diseases including hypertension and diabetes. The microvascular system is of primary involvement in such conditions, and the retina is the only anatomical site where the microvasculature can be directly observed. The objective assessment of retinal vessels has long been considered a surrogate biomarker for systemic vascular diseases, and with recent advancements in retinal imaging and computer vision technologies, this topic has become the subject of renewed attention. In this paper, we present a novel dataset, dubbed RAVIR, for the semantic segmentation of Retinal Arteries and Veins in Infrared Reflectance (IR) imaging. It enables the creation of deep learning-based models that distinguish extracted vessel type without extensive post-processing. We propose a novel deep learning-based  methodology, denoted as SegRAVIR, for the semantic segmentation of retinal arteries and veins and the quantitative measurement of the widths of segmented vessels. Our extensive experiments validate the effectiveness of SegRAVIR and demonstrate its superior performance in comparison to state-of-the-art models. Additionally, we propose a knowledge distillation framework for the domain adaptation of RAVIR pretrained networks on color images. We demonstrate that our pretraining procedure yields new state-of-the-art benchmarks on the DRIVE, STARE, and CHASE\_DB1 datasets.\\
Dataset link: \href{https://ravirdataset.github.io/data}{https://ravirdataset.github.io/data}
\end{abstract}

\begin{IEEEkeywords}
Retinal Image Analysis, Deep Learning, Semantic Segmentation, Vascular Width Estimation, Ophthalmology.
\end{IEEEkeywords}

\section{Introduction}
\label{sec:intro}

The retina and its vasculature are directly visible due to the optically clear media of the human eye; thus, the ability to image the retina rapidly and non-invasively using a variety of modalities provides a window into vital organs such as the brain, heart, and kidneys. Analysis of the retinal vasculature offers a unique avenue to monitor and assess changes associated with systemic diseases such as hypertension, diabetes, and neurodegenerative disorders. The sequelae of these conditions, including stroke, coronary heart disease, and dementia, represent major causes of morbidity and mortality in the developed world~\cite{tekin2004relationship, ong2013hypertensive, omotoso2016relationship}. To date, most classification schemes for retinal vascular changes due to these conditions, particularly in the early stages of disease, depend on the human assessment of qualitative morphological changes~\cite{downie2013hypertensive, cuspidi2015updated}.

\begin{figure}
\centering
{\includegraphics[width=0.32\linewidth,height=0.32\linewidth]{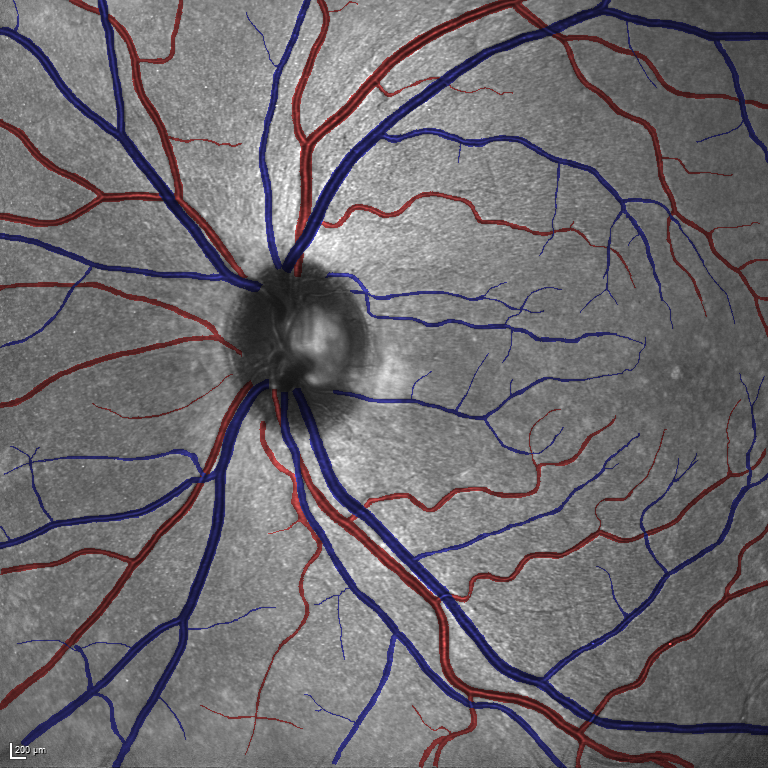}}
\hfill
{\includegraphics[width=0.32\linewidth,height=0.32\linewidth]{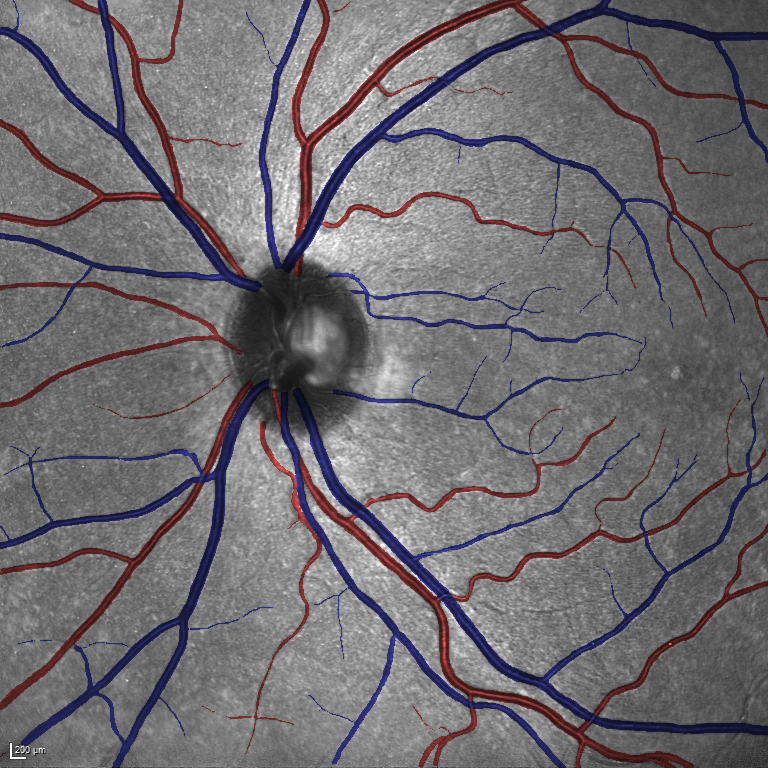}}
\hfill
{\includegraphics[width=0.32\linewidth,height=0.32\linewidth]{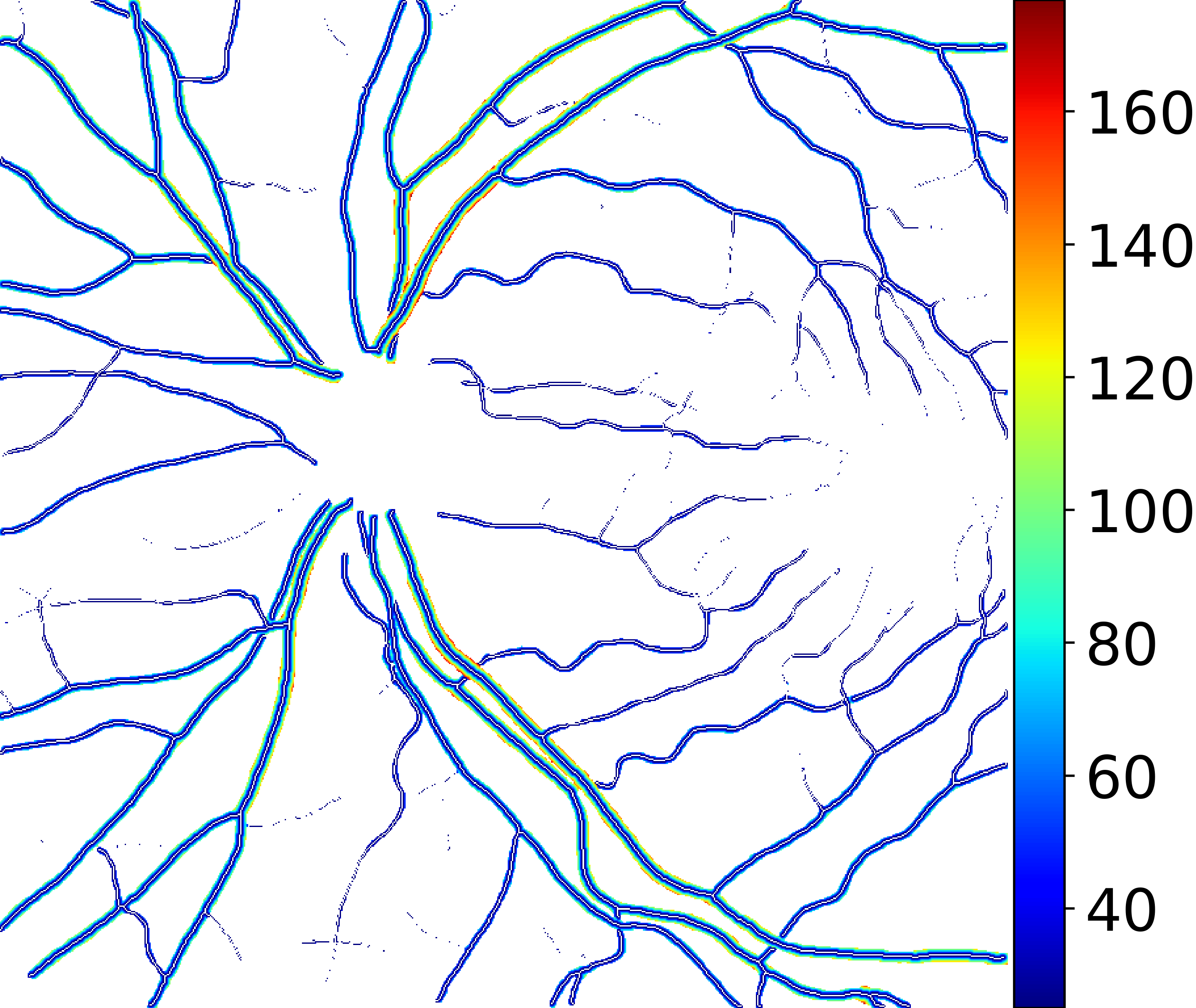}}
\hfill \\[6pt]
{\includegraphics[width=0.32\linewidth,height=0.32\linewidth]{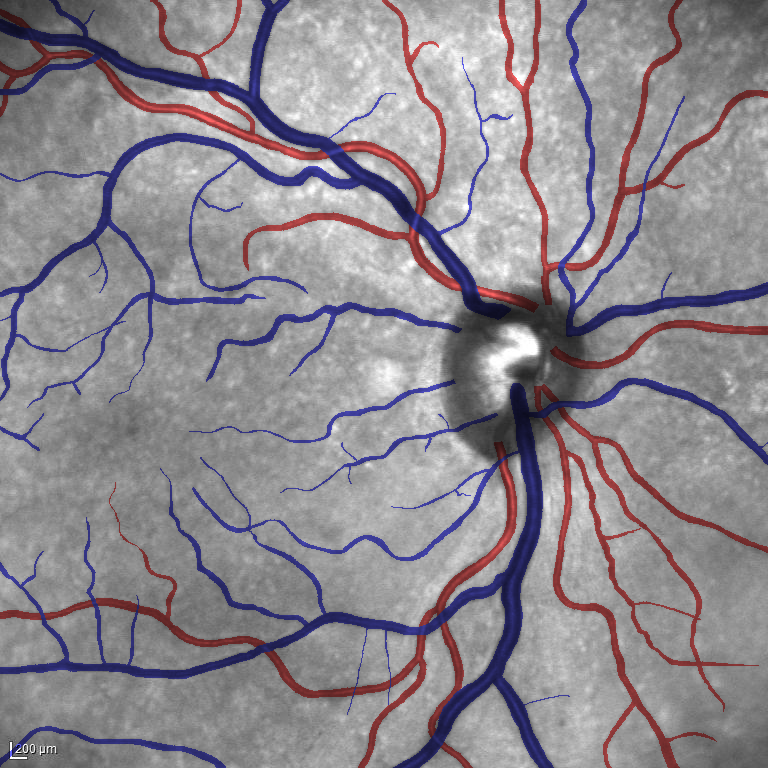}}
\hfill
{\includegraphics[width=0.32\linewidth,height=0.32\linewidth]{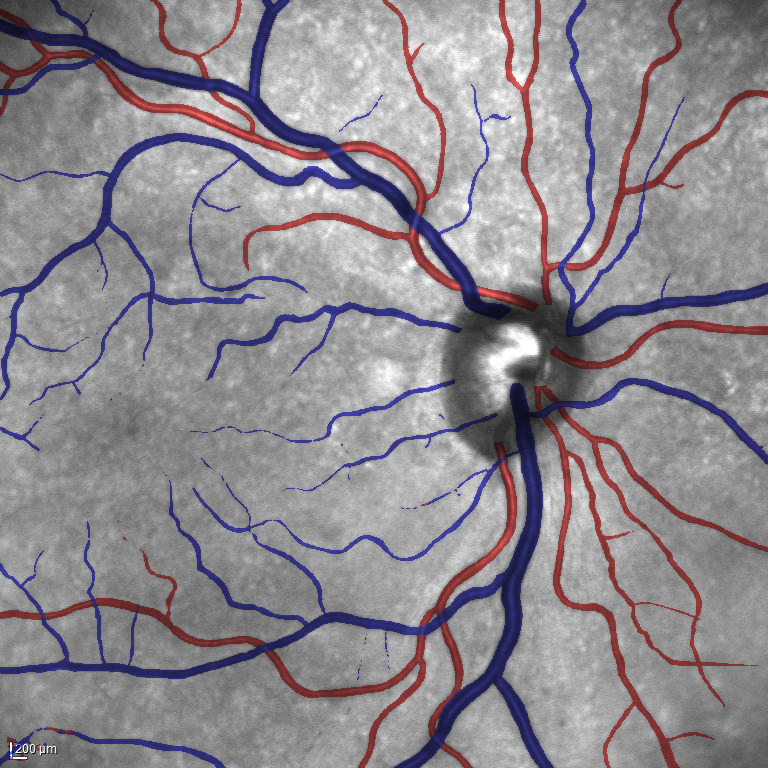}}
\hfill
{\includegraphics[width=0.32\linewidth,height=0.32\linewidth]{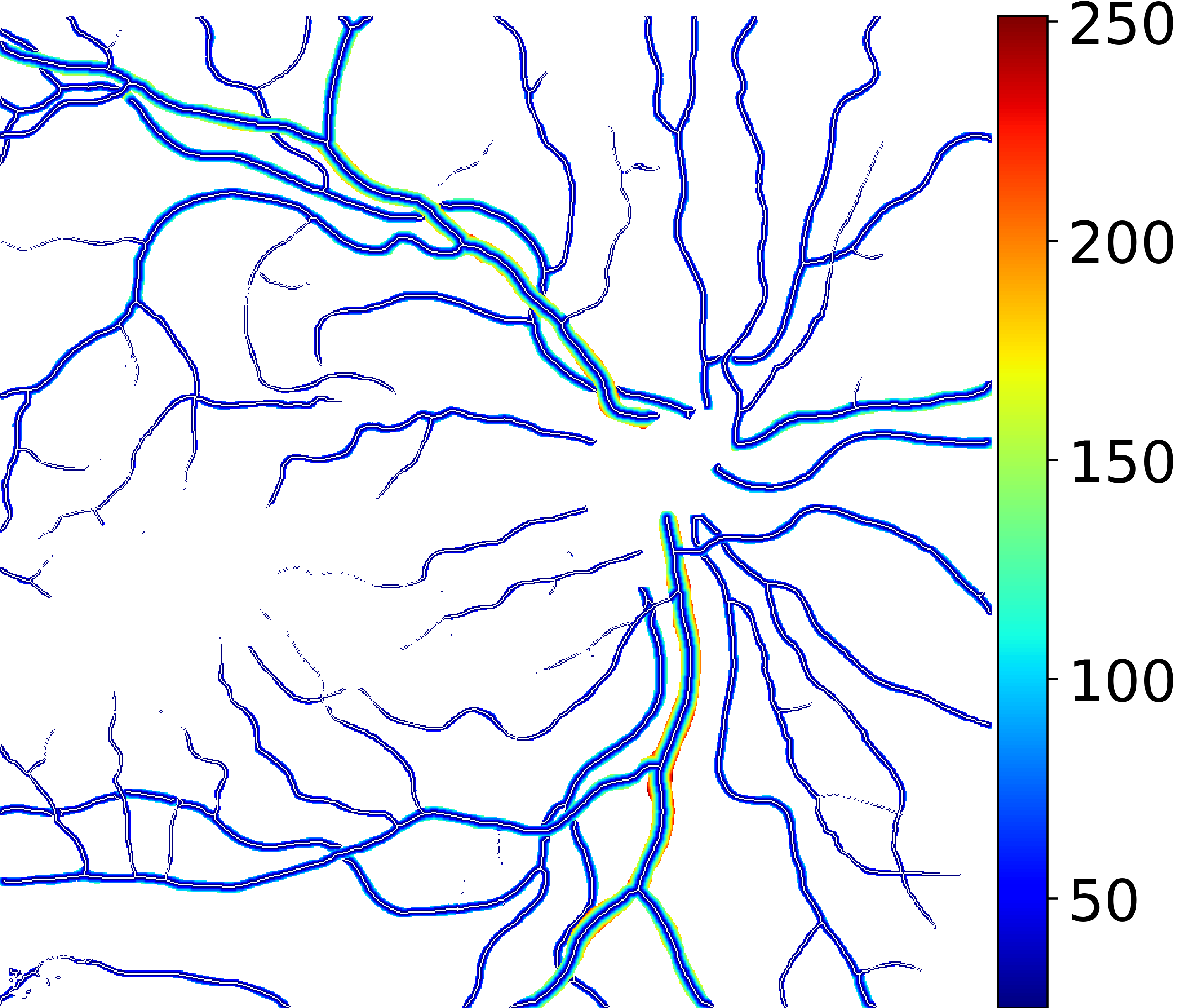}}\\[6pt]

\makebox[0.32\linewidth]{(a) Reference}
\makebox[0.32\linewidth]{(b) Output} 
\makebox[0.32\linewidth]{(c) Width Map}
\caption[Example IR images]{The RAVIR dataset enables semantic understanding of retinal vessels in IR imaging. (a) Reference labels. (b) SegRAVIR semantic segmentation output. (c) Pixel-wise vessel width measurement. Red and blue masks denote artery and vein classes, respectively. Width measurement map presented in microns.}
\label{fig:introfig}
\end{figure}

We and others hypothesize that the assessment of retinal vasculature changes can elicit biomarkers associated with early, even asymptomatic, disease states~\cite{guo2020association}. For example, hypertension is known to cause structural changes in the macro and micro vasculature of vital organs throughout the body, including the brain, heart, and kidneys. As retinal vasculature is both functionally and pathologically linked to cerebral vasculature, it exhibits similar morphological changes~\cite{patton2005retinal}. However, early structural changes to the vasculature are inconspicuous, and detection and analysis of these early changes is often difficult due to the non-quantitative modalities used in routine clinical practice, including direct examination, fundus photography, and angiography.

The accessibility of the retinal vasculature to multimodal imaging provides the opportunity to quantitatively assess prognosis, risk, and response to treatment. Infrared Reflectance (IR) imaging is regarded as an ideal modality to evaluate the retinal vasculature due to several advantages including the ability to capture images through a non-dilated pupil, increased patient comfort due to lack of visible light flash, and better penetration through opaque media. Additionally, previous studies have found a substantial correlation in image resolution and utility between IR and color imaging modalities~\cite{ajaz2019relation}.

While systemic etiologies affect both arterial and venous vasculature, the morphological changes between the two may differ significantly. In the early stages of hypertensive retinopathy, for example, increased blood pressure triggers autoregulatory mechanisms that result in diffuse narrowing of the arterial vessels. If the elevated pressure persists, a compensatory thickening of the arterial walls occurs, which can be clinically observed by the resultant widening and accentuation of the vascular light reflex (``copper wiring'' or ``silver wiring''). By contrast, physiologic changes in venous vasculature include dilated and tortuous venules resulting from venous compression at arteriovenous crossing points subsequent to thickening of the adjacent arterial walls. Additionally, these physiological changes progress in a well-understood and sequential manner that is correlated to the duration and severity of disease. For this reason, clinically observable changes in retinal vasculature are among the primary metrics used in the diagnosis, progression, and response to treatment.

The development of an automated, reliable tool that reproducibly quantifies changes in the retinal vasculature in response to disease and intervention promises to augment or even disrupt current evaluation and treatment paradigms. By enabling physicians to detect disease, predict outcomes, and assess interventions at an earlier stage, such a tool can serve in the management of systemic conditions such as hypertension and diabetes with the potential to dramatically improve patient outcomes and public health at large. 

A critical step in tracking such important structural changes of the retinal vasculature is segmentation of the retinal vessels in ophthalmologic images, as it enables localization of the veins and arteries and the extraction of relevant information such as the change in diameters of the vessels. The automated segmentation and serial analysis of both arterial and venous retinal vasculature is therefore a potentially valuable capability. Since the advent of ``deep learning'' in the field of computer vision, Convolutional Neural Networks (CNNs) have been widely employed for the automated segmentation of medical images, including retinal images. Using stochastic gradient descent optimization approaches, CNNs can be trained end-to-end to automatically learn and extract important features. However, training requires annotated datasets with detailed, pixel-wise ground truth labels, and such annotation processes are often notoriously time-consuming due to the complex structure of retinal vessels.\footnote{Self-supervised and unsupervised deep learning approaches do not require annotations, but supervised approaches usually perform significantly better.}
Consequently, publicly available annotated datasets for retinal vessel segmentation tend to be relatively small in size, as is often the case in medical imaging, a domain that is infamously deficient in adequate datasets.

In this paper, we introduce a dataset for training deep learning CNN models to perform detailed, pixel-resolution ``semantic segmentation'' of Retinal Arteries and Veins in IR imaging, dubbed RAVIR. In conjunction with our new RAVIR dataset, we propose a novel deep learning-based methodology, denoted SegRAVIR, which is tailored to the semantic segmentation of retinal arteries and veins. Specifically, SegRAVIR features a two-stream encoder-decoder CNN architecture consisting of main and auxiliary streams with shared encoders to simultaneously segment the vessels and reconstruct the input images. The role of the auxiliary stream is mainly to regularize the main stream and help the shared encoder learn effective representations of the retinal vessels.  

Since the RAVIR dataset provides precise pixel-scale ground-truth information, we can leverage the output segmentation masks to measure the diameters of retinal arteries and veins. This can enable monitoring the morphological changes of retinal vasculature in a quantitative manner. Our extensive experiments validate the effectiveness of SegRAVIR, as it proves to outperform existing methodologies by a considerable margin. Furthermore, when trained from scratch on another color retinal artery and vein segmentation dataset called RITE, SegRAVIR again achieves state-of-the-art performance.

Additionally, we propose a novel knowledge distillation framework for the domain adaptation from IR to color images of models pretrained on the RAVIR dataset. We demonstrate the effectiveness of our domain adaptation approach by fine-tuning and testing SegRAVIR on three popular color retinal vessel segmentation datasets, DRIVE, STARE, and CHASE\_DB1.   

To summarize, our specific contributions are as follows: 
\begin{enumerate}
   \item We introduce RAVIR, a novel dataset for semantic segmentation and quantitative analysis of retinal arteries and veins in the IR imaging modality. 
   \item We propose a novel framework with tailored loss functions for the segmentation task. To this end, we propose the proxy task of reconstructing input images by using an autoencoder to regularize the segmentation output and improve its fine-grained details.
   \item We provide extensive quantitative benchmarks for performance comparison of our RAVIR framework against state-of-the-art methodologies for retinal vessel segmentation. 
   \item We propose a novel knowledge distillation framework for the domain adaptation to color images of models pretrained on RAVIR, and our results establish new state-of-the-art benchmarks on the publicly available DRIVE, STARE, and CHASE\_DB1 datasets by a substantial margin.

\end{enumerate}

\begin{table*} \centering
	\caption{Comparison between the RAVIR dataset and other publicly available} retinal vessel segmentation datasets
	\label{table:datasetcompare}
	\resizebox{\linewidth}{!}{
	\begin{tabular}{lccccclc}
	  \toprule
		Dataset &Resolution&FOV&Train Size&Test Size&AV Label&Camera&Modality \\ 
		\midrule
		\textbf{RAVIR} &$768\times768$&30&26&20&Yes&Heidelberg
        Spectralis&IR \\
        STARE~\cite{hoover2000locating} &$605\times700$&35&20&-&No&Topcon TRV-50&Color   \\
		DRIVE~\cite{staal2004ridge} &$565\times584$&45&20&20&Yes&Canon CR-5 non-mydriatic&Color    \\
        RITE~\cite{hu2013automated}
        & $565\times584$ & 45 & 20 & 20 & Yes & Canon CR-5 non-mydriatic & Color\\
		CHASE\_DB1~\cite{fraz2012ensemble} &$999\times960$&30&20&8&No&Nidek NM-200-D&Color   \\
		IOSTAR~\cite{zhang2016robust} &$1024\times 1024$&45&30&-&Yes&EasyScan&Color   \\
		LES-AV~\cite{orlando2018towards} &$1620\times1444$&30&22&-&Yes&Canon CR-1&Color    \\
		HRF~\cite{budai2013robust} &$3504\times2336$&45&45&-&Yes&Canon CR-1&Color   \\
		ROSE~\cite{ma2020rose} &$304\times 304$&-&90&27&No&RTVue-XR&OCTA   \\
		RECOVERY-FA19~\cite{ding2020novel} &$3900\times 3072$&200&8&-&No&Optos 200Tx&FA   \\
		PRIME-FP20~\cite{ding2020weakly} &$4000\times 4000$&200&15&-&No&Optos 200Tx&FA   \\
		\bottomrule
	\end{tabular}
	}
\end{table*}

\section{Related Work}
\label{sec:relatedwork}

In this section, we review and discuss the publicly available datasets for training models to perform retinal artery and vein segmentation in different imaging modalities (Table~\ref{table:datasetcompare}), as well as the notable deep-learning-based methodologies underlying such segmentation models.

\subsection{Retinal Artery and Vein Segmentation Datasets}
\label{sec:datasets}

Traditional color photographs have long been utilized in the screening evaluation of retinal vascular diseases.

One such early effort is the STructured Analysis of the REtina (STARE) dataset~\cite{hoover2000locating}, which was created at the University of California, San Diego, with the aim of developing a system capable of the automated diagnosis of various diseases in the human eye. This dataset comprises 20 color retinal images, acquired by a Topcon TRV-50 camera at a $35^\circ$ field of view (FOV), and sized at $605\times 700$ pixels with 8 bits per RGB color channel. Half of the images in this dataset depict pathology that manifests an abnormal appearance of retinal blood vessels and obscures them.

Another important effort is the Digital Retinal Images for Vessel Extraction (DRIVE) dataset~\cite{staal2004ridge}, which was created as part of a diabetic retinopathy screening program in The Netherlands. It consists of 40 color retinal images with corresponding binary segmentation masks, which are equally divided into train and test sets. In the DRIVE dataset, 7 cases show signs of mild early diabetic retinopathy or conditions such as pigment epithelium changes, pigmentary epithelial atrophy, butterfly maculopathy with pigmented scar in fovea, choroidiopathy, atrophy around the optic disc, background diabetic retinopathy, and other vascular abnormalities. The images were acquired using a Canon CR5 non-mydriatic 3CCD camera with a $45^\circ$ FOV and are sized at $768\times 584$ pixels with 8 bits per RGB color channel.

Using the original images in the DRIVE dataset, the Retinal Images vessel Tree Extraction (RITE) dataset~\cite{hu2013automated} was developed to support the segmentation and classification of retinal arteries and veins. 
Qureshi \textit{et al.}~\cite{qureshi2013manually} also presented a manually labeled dataset with pixel-wise annotations for retinal arteries and veins for images in the DRIVE dataset.

The Child Heart and Health Study (CHASE) in England yielded the CHASE\_DB1 dataset~\cite{fraz2012ensemble}, including 28 color retinal images from both eyes of 14 child subjects. The images were captured by a hand-held Nidek NM-200-D fundus camera with a resolution of $1280\times 960$ pixels at a $30^\circ$ FOV. The vessels in these images are poorly contrasted against the background. Additionally, the artery cross-sectional intensity depicts a central vessel reflex. Eight images are used as the training set and the remainder as a test set.

The IOSTAR~\cite{zhang2016robust} and LES-AV~\cite{orlando2018towards} datasets contain color retinal images that include artery and vein pixel-wise annotations. The IOSTAR dataset consists of 30 images of size $1024\times 1024$ pixels acquired through a Scanning Laser Ophthalmoscopy (SLO) methodology from patients suffering from diabetic retinopathy, whereas the LES-AV dataset has 22 images of glaucomatous patients.

Another commonly used dataset is the High Resolution Fundus (HRF) dataset~\cite{budai2013robust}, which contains 45 color retinal images of size $3504\times 2336$ pixels captured by a Canon CR-1 fundus camera with a $45^\circ$ FOV. The dataset includes images of 15 subjects with diabetic retinopathy, 15 subjects with glaucoma, and 15 healthy subjects.

Aside from IR imaging and RAVIR, and in addition to the aforecited color image datasets, retinal vessel segmentation datasets have recently been curated in other imaging modalities. The ROSE dataset~\cite{ma2020rose} provides 117 Optical Coherence Tomography Angiography (OCTA) images of $304\times 304$ pixel resolution from 39 subjects suffering from Alzheimer's disease and, with regard to the Fluorescein Angiography (FA) modality, the RECOVERY-FA19 dataset~\cite{ding2020novel} contains 8 ultra-wide-field images of size $3900\times 3072$ pixels from patients participating in the Intravitreal Aflibercept for Retinal Non-Perfusion in Proliferative Diabetic Retinopathy trial~\cite{wykoff2019intravitreal}. Another FA dataset is the PRIME-FP20 dataset~\cite{ding2020weakly}, comprising 15 images with a resolution of $4000\times 4000$ pixels acquired from patients with diabetic retinopathy.

\subsection{Retinal Vessel Segmentation Methodologies}

Subsequent to the success in medical and biomedical image segmentation of the seminal U-Net~\cite{ronneberger2015u} architecture, many efforts have been proposed to leverage its encoder-decoder paradigm in the context of retinal vessel segmentation. Zhuang \textit{et al.}~\cite{zhuang2018laddernet} proposed an architecture based on U-Net that utilizes multiple-path networks to leverage a path-wise formulation in segmenting retinal vessels. Conversely, Wang \textit{et al.}~\cite{wang2019dense} proposed a Dense U-Net architecture with a patch-based training strategy and extensive data augmentation, which outperforms the baseline U-Net by a small margin. Jiang \textit{et al.}~\cite{jiang2019automatic} developed a multi-path encoder-decoder architecture that fused the outputs of the network at different scales to obtain the final predictions. 

Furthermore, Alom \textit{et al.}~\cite{alom2019recurrent} introduced a recurrent residual U-Net architecture in which residual convolutional layers are used for optimal feature representation learning. Jin \textit{et al.}~\cite{jin2019dunet} proposed a deformable U-Net architecture that captures various scales and shapes of the vessels by adjusting the adaptive receptive field of the deformable convolutional layers. Wu \textit{et al.}~\cite{wu2019vessel} used a U-like encoder-decoder architecture that leverages Inception residual blocks for improved feature representation and is supervised in a multi-scale manner to preserve features of different sizes. Zhang~\textit{et al.}~\cite{zhang2019attention} proposed an attention-guided network that effectively captures vessel structural information. Wang \textit{et al.}~\cite{wang2019dual} utilized two encoders, in a U-Net like architecture, to preserve the spatial information and semantic representations separately. Gu~\textit{et al.}~\cite{gu2019net} proposed a context encoder network, CE-Net, to effectively capture high-level semantic and spatial content for use in segmenting vessels. Additionally, Zhang~\textit{et al.}~\cite{zhang2019net} proposed an edge-attention guidance network to improve the accuracies of the boundaries of segmented vessels.  

Recently gaining traction is the application of cascaded architectures where several encoder-decoders are chained together to refine the segmentation outputs. Although these approaches typically result in more refined boundaries, they are memory-intensive and the inference time is substantially increased due to the complexity of the model. Yan \textit{et al.}~\cite{yan2018three} introduced a methodology in which a three-step CNN was employed to separately segment thick and thin vessels and learn how to effectively fuse the results and refine the predictions. Francia \textit{et al.}~\cite{francia2020chaining} proposed to use an architecture consisting of two residual U-Nets in which the second refines the feature maps of the first and resolves the ambiguities. Li \textit{et al.}~\cite{Li_2020_WACV} used a cascade of U-Net architectures that share their weights and are connected by mid-level skip connections, thus yielding a deeper CNN that can effectively disambiguate the uncertain regions in the output segmented vessels. Guo \textit{et al.}~\cite{guo2020bscn} introduced a bidirectional symmetric cascade with adaptive supervision of each layer according to the diameter scale of each vessel. 

In addition to graph-based methodologies~\cite{dashtbozorg2013automatic,estrada2015retinal,hu2015automated}, several efforts have applied deep learning-based methods to the semantic segmentation of arteries and veins as distinct vessels in color retinal images. Welikala \textit{et al.}~\cite{welikala2017automated} used an architecture consisting of three convolutional layers and three fully-connected layers to segment vessels in small $25\times 25$ image patches. Hemelings \textit{et al.}~\cite{hemelings2019artery} proposed another patch-based approach to vessel segmentation by leveraging a regular U-Net with categorical cross-entropy as the loss function. Girard \textit{et al.}~\cite{girard2019joint} introduced a whole-image approach with a U-Net-like encoder-decoder in which the output predictions of the CNN are propagated through a graph representation of the retinal vasculature based on a likelihood score for detecting arteries, but the model requires training on small patches that are centered on retinal vessels, which is a limitation relative to the method by Hemelings \textit{et al.}~\cite{hemelings2019artery}; however, both methods require ad-hoc image pre-processing techniques to achieve optimal performance. Xu \textit{et al.}~\cite{xu2018simultaneous} proposed a related whole-image approach requiring extensive data augmentation, and the final segmentation maps are obtained via a heuristic for assigning pixel-wise outputs. Kang \textit{et al.}~\cite{kang2020avnet} introduced an architecture for segmenting retinal arteries and veins consisting of a GoogLeNet~\cite{szegedy2015going} encoder, a decoder, and a weighted-attention fusion module that predicts the final segmentation outputs by fusing and refining segmentation probabilities of each semantic class; however, the absence of high-resolution skip connections between the encoder and decoder limits the ability to exploit multi-scale representations. Ma \textit{et al.}~\cite{ma2019multi} introduced a model consisting of a pretrained U-Net incorporating a ResNet~\cite{he2016deep} encoder with two parallel branches for extracting common retinal artery and vein representations, focusing on discriminative features whose segmentation output is generated by concatenating the outputs of each branch and using a customized activation block that emphasizes the importance of capillary vessels. Chen \textit{et al.}~\cite{chen2020tr} proposed a  Generative Adversarial Network (GAN) approach to improving the connectivity of segmented retinal arteries and veins. Morano \textit{et al.}~\cite{morano2021simultaneous} described a U-like architecture to independently segment retinal arteries, veins, and vessels using Binary Cross Entropy (BCE) as the loss function. Despite these advances, a primary shortcoming of the aforementioned methodologies is frequent classification errors in the detected vessels.

\section{The RAVIR Dataset}
\label{sec:ravirdataset}

Recently there has been a substantial increase in the utilization of single wavelength confocal imaging that only captures reflected light passing through a pinhole, which will allow capturing high quality, high contrast fundus images. The images in our RAVIR dataset were captured using infrared (815nm) Scanning Laser Ophthalmoscopy (SLO), which in addition to having higher quality and contrast, is more convenient for the patient and is less affected by opacities in optical media and pupil size. This imaging modality is also commonly used in ophthalmology clinics. The RAVIR dataset consists of 46 IR retinal images from the UCLA Stein Eye Institute imaging database, divided into train and test sets of 26 and 20 images, respectively. We used 4 images from the 26 test images as our validation set. The images were captured using a Heidelberg Spectralis camera with a $30^\circ$ FOV. They are sized at $768\times 768$ pixels and compressed in the Portable Network Graphics (PNG) format. Each pixel in the images has a reference length of 12.5 microns. Table~\ref{table:datasetcompare} compares our RAVIR dataset against other retinal vessel segmentation datasets.

Our study was carried out with the approval of the Institutional Review Board at UCLA and adhered to the tenets of the Declaration of Helsinki. Table~\ref{table:avir1} reports the demographics of the study population. The average age of the subjects was 55 years, with a range from 19 to 88. There were 29 male and 17 female subjects. One eye from each patient was imaged for the dataset, and there were 24 right eyes and 22 left eyes. Twenty-three images were normal, while 23 images depicted retinal pathology. The train and test sets of the RAVIR dataset have an equal number of healthy and pathological cases.

\begin{table} \centering
	\caption{RAVIR dataset demographics}
	\label{table:avir1}
	\begin{tabular}{ll}
	  \toprule
		Statistics  &Quantity    \\ \midrule
		Average Age (years) &$55\pm18$  \\
		Age Range (years) & $19-88$  \\
		Number of Males & $29$  \\
		Number of Females & $17$  \\
		Number of Right Eyes &$24$   \\
		Number of Left Eyes &$22$   \\
		\bottomrule
	\end{tabular}
\end{table}

\begin{table} \centering
	\caption{IR imaging manifestations of study population in RAVIR}
	\label{table:avir2}
	\setlength{\tabcolsep}{0.8mm}
	\begin{tabular}{lcl}
	 \toprule
		Clinical Finding  & \#Eyes & Imaging Manifestation  \\ \midrule
		Retinal vein occlusion &$10$ &Venous tortuosity, retinal hemorrhage\\
		Hypertensive retinopathy &$10$&Vessel narrowing, arteriovenous nicking  \\
		Peripapillary Atrophy & $8$&Increased background reflectivity\\
		Diabetic Retinopathy &$3$&Intraretinal hemorrhage, vessel sclerosis   \\
		Isolated vessel tortuosity &$3$&Abnormal vessel tortuosity   \\
		High Myopia &$2$&Transmission of choroidal vessels   \\
		Media opacities &$1$& Retinal haziness  \\
		\bottomrule
	\end{tabular}
\end{table}

\begin{figure*} \centering
\includegraphics[width=\textwidth]{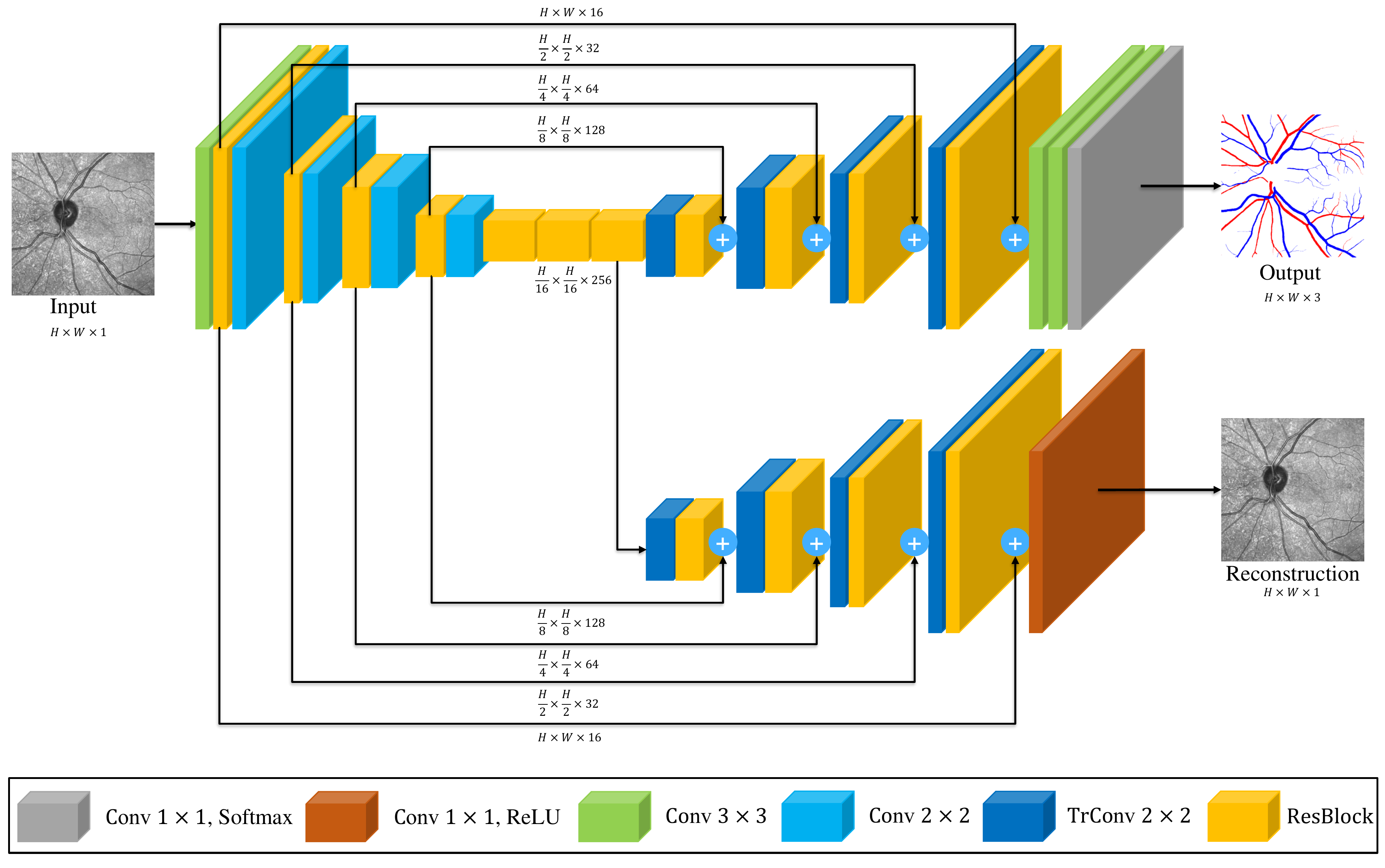}
\caption{SegRAVIR consists of a main stream for segmenting the arteries and veins as well as an autoencoder that learns to reconstruct the input and acts as a regularizer. The architectural building blocks are depicted at bottom ---Conv, TrConv, and ResBlock denote convolution, transposed convolution, and residual blocks, respectively.}
\label{fig:pipeline_avir}
\end{figure*}

As described in Table~\ref{table:avir2}, a wide variety of retinal pathologies are covered by the dataset. These conditions may all affect the retinal vascular pattern and infrared retinal background. For example, diabetic retinopathy and vein occlusions may cause vascular caliber changes and signal blocking from retinal hemorrhage or edema. Peripapillary atrophy causes increased infrared reflectance around the Optic Nerve Head (ONH). Hypertensive retinopathy can manifest as vessel narrowing and arteriovenous nicking, which is the segmental narrowing of veins at arterial crossing points. Severe myopia can cause increased signal transmission of choroidal vessels. 

Manual pixel-wise annotations were performed and verified by our experienced retinal image analysis specialist. Per-pixel labels were applied to all vessel regions that could accurately be identified as artery or vein. Vessels were labeled over the ONH in those images where the arteries and veins could be resolved. However, for images in which the veins and arteries over the ONH were indistinguishable, the ONH region was blocked for masking.  

\section{SegRAVIR Methodology}
\label{sec:ravirmethod}

This section develops our SegRAVIR methodology, including the network architecture and loss function, as well as our vessel width estimation technique and domain adaptation strategy.

\subsection{The SegRAVIR Architecture}
\label{sec:ravirarch}

To analyze retinal arteries and veins, we devise an encoder-decoder architecture dubbed SegRAVIR (Fig.~\ref{fig:pipeline_avir}), which consists of a main stream for segmenting and an auxiliary stream for reconstructing the input images. The main stream leverages residual blocks and skip connections that connect the encoder and decoder. The auxiliary stream benefits from a decoder that learns how to reconstruct the input images and acts as a regularizer. SegRAVIR's building blocks are residual blocks consisting of two $3\times3$ convolutional layers and an identity skip connection that adds the input of the block to the convolutional outputs. Each convolutional layer with kernel $W$ is followed by a Rectified Linear Unit (ReLU) and a batch normalization $\textit{BN}_{\gamma,\beta}$ with learnable parameters $\gamma$ and $\beta$:
\begin{equation}
C(i,j) = \textit{BN}_{\gamma,\beta}\left(\textrm{Re}\left(\sum_{t=-1}^{+1}\sum_{l=-1}^{+1} K[t,l] W[i-t,j-l]\right)\right),
\label{eq:conv}
\end{equation}
where $C$ and $K$ denote the input and output of the convolution layer, respectively, and $i$, $j$, $t$, $l$ are pixel indices. Additionally, the output of each convolution is fed into a dropout layer for regularization. SegRAVIR's encoder spans across 4 different resolutions, each of which consists of a residual block followed by a convolutional layer with a stride of 2 and appropriate padding to reduce the resolution of learned representations by a factor of 2. At its lowest resolution, the output of the convolution layer is fed into 3 consecutive residual blocks and then into the decoder. 

SegRAVIR's decoder has 4 corresponding resolutions consisting of transposed convolutional layers with a stride of 2 and appropriate padding to increase the resolution of learned representations by a factor of 2, followed by a residual block. At each resolution, the output of the residual blocks in the encoder is added to the output of the corresponding residual block in the decoder via a skip connection. Finally, the output of the decoder is fed into two consecutive $3\times3$ convolutional layers with ReLU activation and batch normalization, followed by a $1\times1$ convolutional layer with softmax activation and three output channels that correspond to background, vein, and artery classes.

Furthermore, we employ an auxiliary stream that learns to reconstruct the input images and acts as a regularizer. It is an auto-encoder that regularizes the shared encoder by learning an additional proxy task of image reconstruction in order to avoid overfitting. At each resolution, the outputs of residual blocks in the main-stream encoder are fed into a decoder that shares the same architecture as the main-stream decoder, except for the last three layers which use a $1\times1$ convolutional layer followed by a ReLU activation function. The output of the auxiliary stream has the same size as the input image; hence, a proper image-based reconstruction loss can be employed at the original resolution. 

Unlike a typical U-Net, the SegRAVIR model leverages residual blocks in both the encoder and decoder of the main stream and benefits from three residual blocks in the bottleneck of the architecture, thus enabling the capture of  fine-grained details of the retinal vasculature. Moreover, its auxiliary stream regularizes the segmentation outputs, further improving performance. Fig.~\ref{fig:pipeline_avir} provides additional SegRAVIR architectural details.

\subsection{SegRAVIR Loss Function}
\label{sec:ravirloss}

To supervise the outputs of the main stream as well as the autoencoder of the auxiliary stream, we employ a hybrid loss function consisting of the following terms:
\begin{equation}
\mathcal{L}= \lambda_{1} \mathcal{L}_\textrm{Dice} + \lambda_{2} \mathcal{L}_\textrm{CE} + \lambda_{3} \mathcal{L}_\textrm{L2},
\label{eq:losstot}
\end{equation}
where $\lambda_{i}$ are weighting parameters associated with the loss terms. Letting the number of pixels and number of semantic classes be $V$ and $K$, respectively, the Dice loss function is defined as
\begin{equation}
\mathcal{L}_\textrm{Dice}= 1-\frac{2}{K}\sum_{k=1}^{K}\frac{\sum_{v=1}^{V} G_{v,k}P_{v,k} }{\sum_{v=1}^{V}G_{v,k}+ \sum_{v=1}^{V}P_{v,k}},
\label{eq:dice1}
\end{equation} 
where $P_{v,k}$ and $G_{v,k}$ denote the output probability prediction and corresponding one-hot encoded ground truth for class $k$ at pixel $v$. The Cross Entropy (CE) loss is defined as
\begin{equation}
\mathcal{L}_\textrm{CE}= -\frac{1}{V}\sum_{v=1}^{V}\sum_{k=1}^{K} G_{v,k}\log P_{v,k},
\label{eq:ce}
\end{equation}
The L2 reconstruction loss of the input image is defined as
\begin{equation}
\mathcal{L}_\textrm{L2}= \lVert X_{i}-X_{r} \rVert_{2}^{2}.
\label{eq:L2}
\end{equation}
where $X_{i}$ and $X_{r}$ denote the input and reconstructed images, respectively. While $\mathcal{L}_\textrm{L2}$ supervises the output of the autoencoder, a combination of $\mathcal{L}_\textrm{Dice}$ and $\mathcal{L}_\textrm{CE}$ serves to supervise the output segmentation of the main stream. The hybrid loss function enables the learning of different semantics as the $\mathbf{\mathcal{L}_\textrm{CE}}$ and $\mathbf{\mathcal{L}_\textrm{Dice}}$ loss terms capture pixel-wise and region-based information, respectively.

\subsection{Vessel Width Estimation}
\label{sec:width}

From the segmentation predicted by the SegRAVIR network, our model measures the diameter of the arteries and veins. To this end, the segmentation probability map is first thresholded, using a constant value of 0.5, to obtain the medial curves of the vessels by iteratively identifying and removing border pixels while maintaining vessel connectivity, in an approach similar to the thinning algorithm presented by Zhang~\textit{et al.}~\cite{Zhang:1984}. Then, the distance transform of the medial curve mask is multiplied with the segmentation mask in a pixel-wise manner. The result is the diameter distance map with respect to the medial curves of the segmented vessels.

\subsection{Knowledge Adaptation via Distillation}
\label{sec:knowledge}

Knowledge distillation~\cite{hinton2015distilling}\cite{adriana2015fitnets}, which was originally proposed as a model compression methodology, enables the training of a smaller student model under the supervision of a larger pretrained teacher network. We leverage knowledge distillation as a feature alignment mechanism for the domain adaptation of RAVIR-pretrained networks to color retinal images. Specifically, we first use transfer learning to create a teacher network in the new modality (color) from a pretrained SegRAVIR model. For this purpose, we use the main stream (encoder and segmentation decoder) of a SegRAVIR network and initialize the weights from a previous pretrained checkpoint. However, we replace the last layer of the network with a $1\times1$ convolutional layer whose weights are randomly initialized, followed by a softmax activation layer to account for the single output channel of the target segmentation task (i.e., binary vessel segmentation). We fine-tune this teacher network on the target image modality. 

\begin{figure} \centering
\includegraphics[width=\columnwidth]{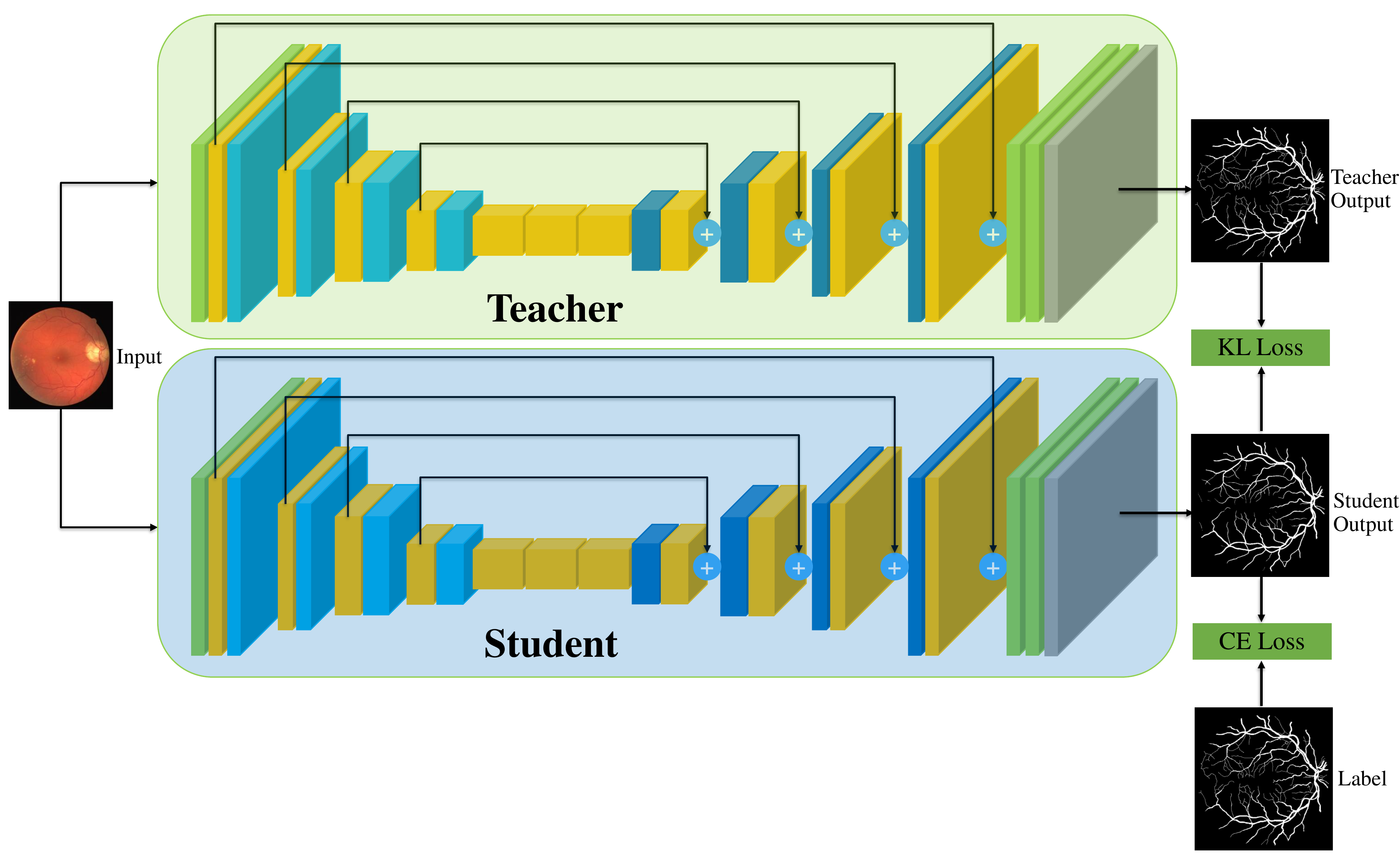}
\caption{Knowledge distillation strategy. A SegRAVIR network fine-tuned on color images supervises the output of a SegRAVIR student network by minimizing the KL divergence of prediction probability distributions.}
\label{fig:pipeline_distill}
\end{figure}

Furthermore, we create a student network with the same architecture and supervise its segmentation outputs according to the soft labels generated by the teacher network predictions. Fig.~\ref{fig:pipeline_distill} overviews the proposed strategy.

Specifically, let $Z_T\in \Re^{H\times W\times 2}$ denote the pixel-wise output logits of the teacher network. We employ temperature $\tau$ scaling to generate soft labels with smoother probability distributions, hence enabling the identification of inter-class relationships, according to
\begin{equation}
Y_{T}=\frac{\exp \left(Z_{T} / \tau\right)}{\sum \exp \left(Z_{T} / \tau\right)}.
\label{eq:softlabel}
\end{equation}
Similarly, we generate soft labels $Y_{S}$ by applying temperature scaling to the output logits $Z_S$ of the student network. Consequently, the distillation loss is calculated as
\begin{equation}
\mathcal{L}_\textrm{Distill} =\lambda_{D}\tau^{2}\mathcal{L}_\textrm{KL}({Y}_{S},{Y}_{T}) + (1-\lambda_{D})\mathcal{L}_\textrm{CE}({Y}_{S},G),
\label{eq:distill}
\end{equation}
where $\lambda_{D}$ is a hyper-parameter, $G$ denotes the segmentation ground truth, and $\mathcal{L}_\textrm{KL}$ denotes the KL divergence loss.

As is shown in Section~\ref{sec:abl1}, our adaptation via distillation strategy is more effective than naive transfer learning, since it addresses the domain gap between the RAVIR dataset and color fundus image datasets by feature alignment in the target space through KL divergence minimization.

\section{Experiments}

\subsection{Implementation Details}
\label{sec:impl}

All the models were implemented in TensorFlow 2.0~\cite{abadi2016tensorflow}.
Experiments were performed on an Nvidia Titan RTX GPU, and an Intel® Core™ i7-7700K CPU @ 4.20\,GHz.

All the networks employ a sliding window approach and were trained on randomly extracted patches of size $256\times256$. Additionally, we used common data augmentation techniques such as random rotations, flipping, and change of contrast in order to increase the available training samples. We trained all the networks for 600 epochs and used the Adam optimization algorithm \cite{kingma2014adam} with minibatches of size 6 and an initial learning rate of 0.001 that is reduced by a factor of 2 every 50 epochs.

In all experiments, the network is evaluated on a validation set consisting of 4 images from the original 26 images, and the instance with the best average F1 score on the artery and vein classes is used for testing.

We used a grid search approach and found the following optimal values for the hyper-parameters in (\ref{eq:losstot}): $\lambda_{1}=\lambda_{2}=1.0$ and $\lambda_{3}=0.001$. These hyper-parameters enable the balancing of their respective loss terms so as to facilitate fast convergence to optimal solutions. In addition, for the hyper-parameters in (\ref{eq:distill}), we use $\tau=3$ and $\lambda_{D}=0.1$. All the baseline methods were implemented by adopting their official released repositories. For all the baseline experiments, we used a combination of Dice and CE as the training loss function. 

\begin{figure*} \centering
\def\x{0.16}
\includegraphics[width=\x\linewidth,height=\x\linewidth]{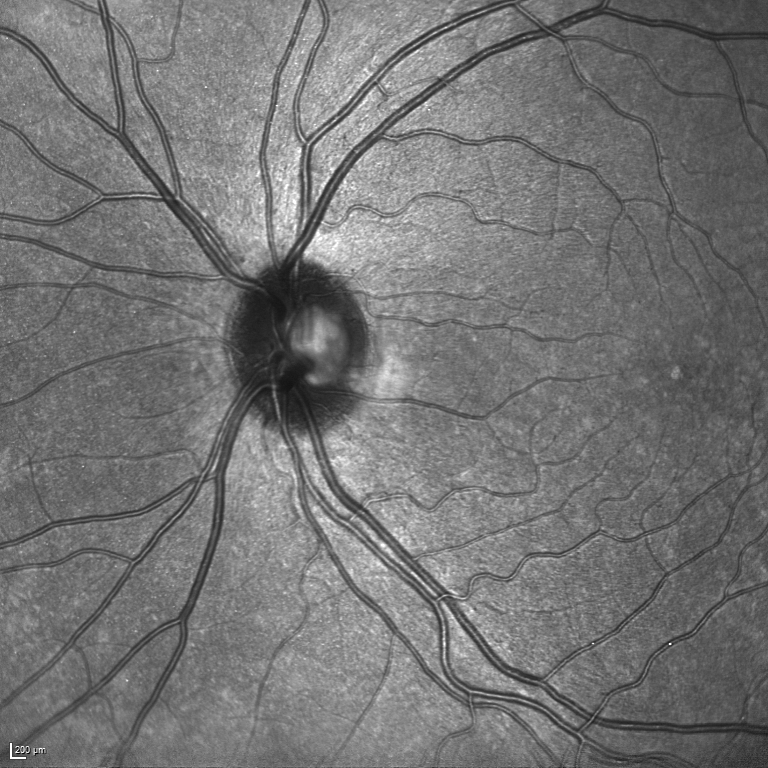}
\hfill
\includegraphics[width=\x\linewidth,height=\x\linewidth]{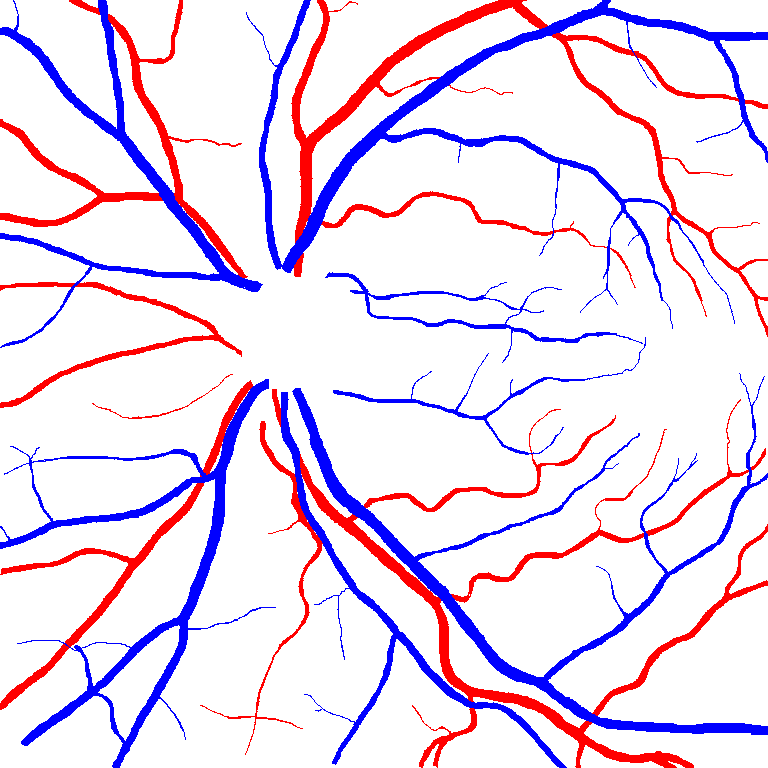}
\hfill
\includegraphics[width=\x\linewidth,height=\x\linewidth]{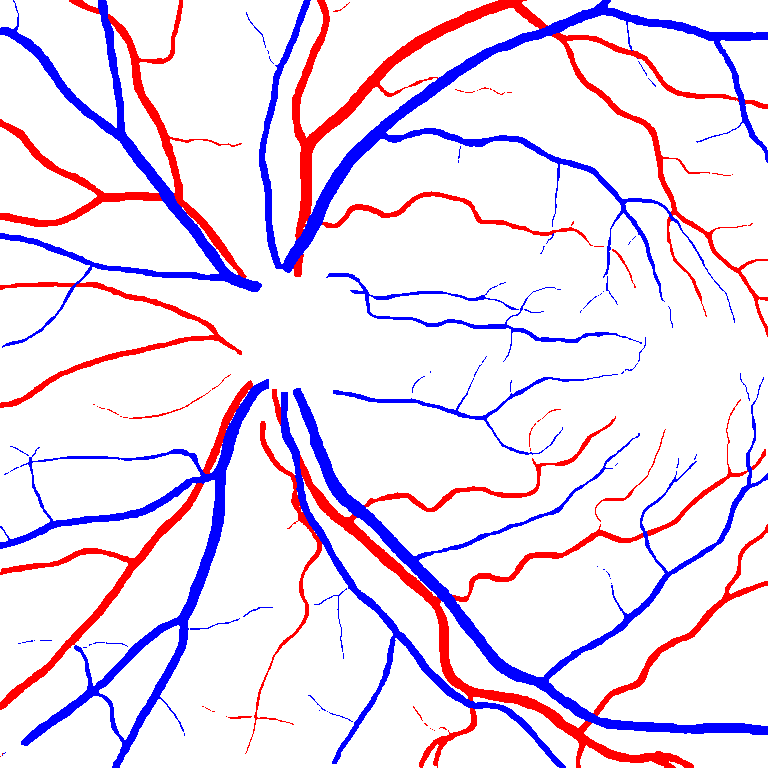}
\hfill
\includegraphics[width=\x\linewidth,height=\x\linewidth]{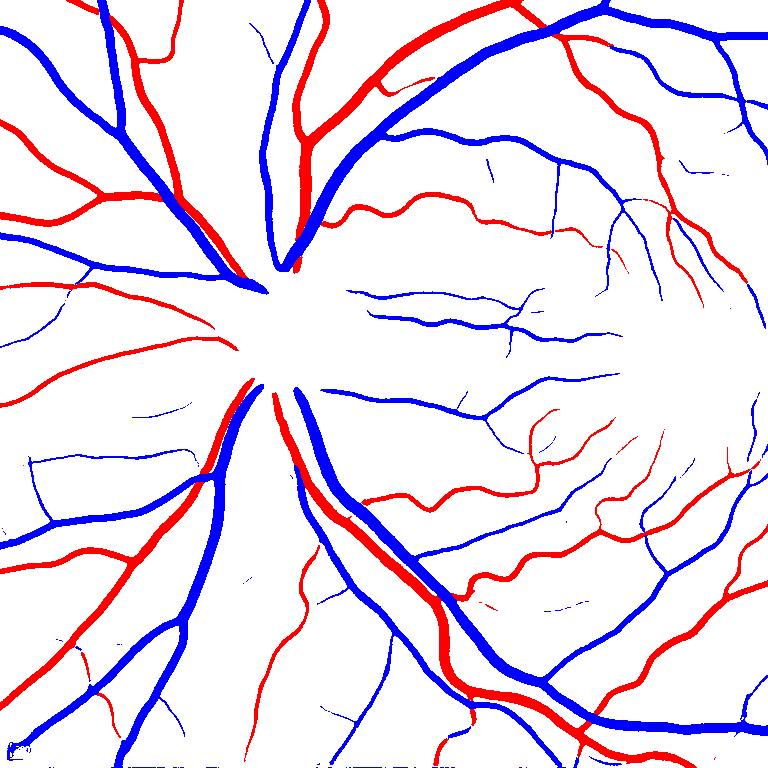}
\hfill
\includegraphics[width=\x\linewidth,height=\x\linewidth]{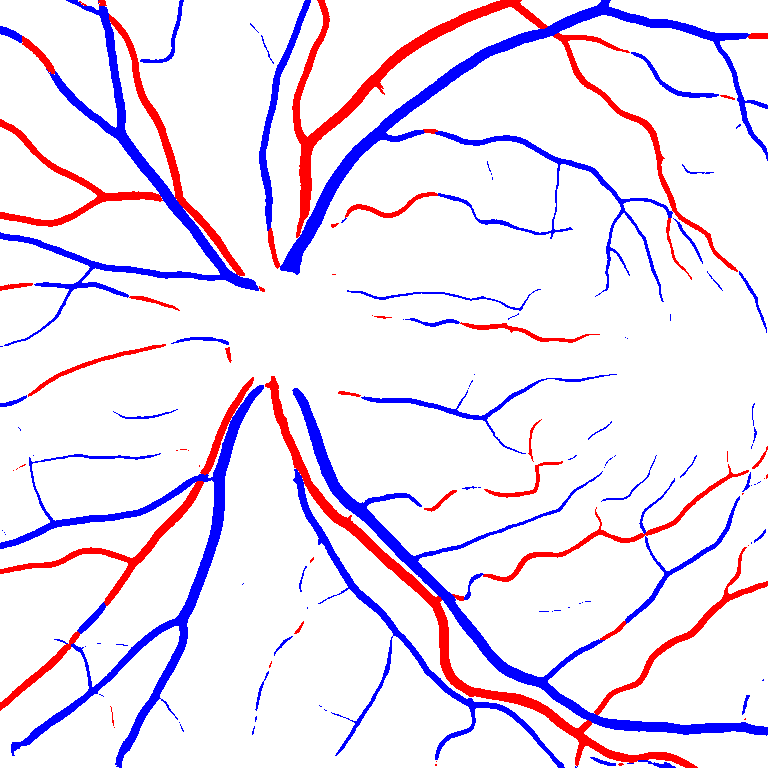}
\hfill
\includegraphics[width=\x\linewidth,height=\x\linewidth]{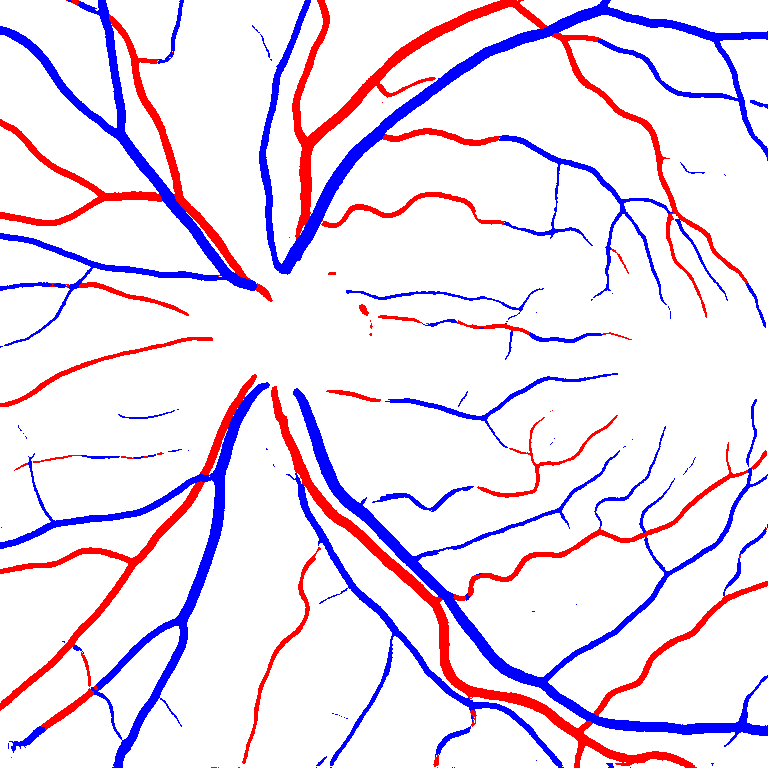}
\\[10pt]
\includegraphics[width=\x\linewidth,height=\x\linewidth]{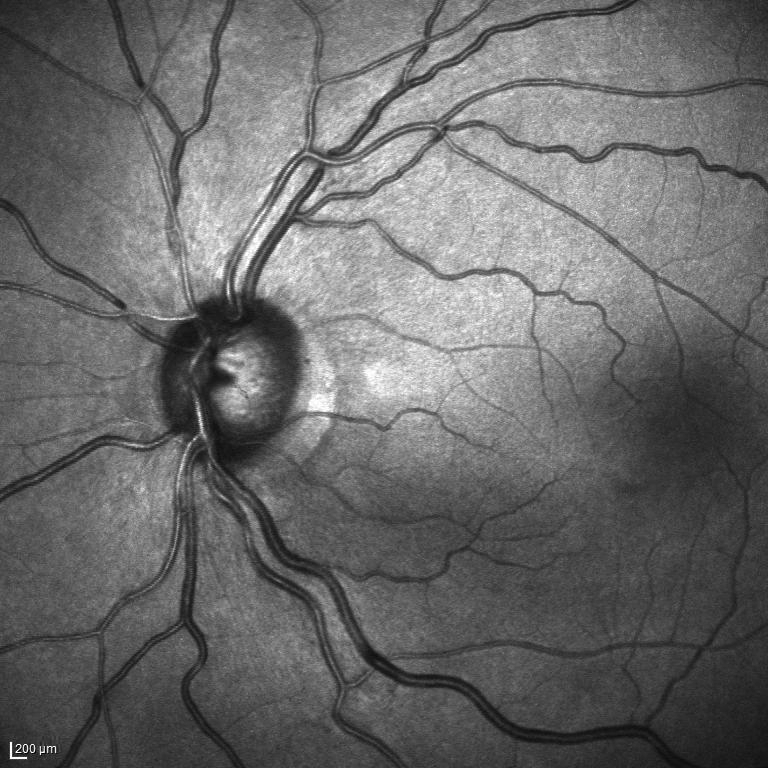}
\hfill
\includegraphics[width=\x\linewidth,height=\x\linewidth]{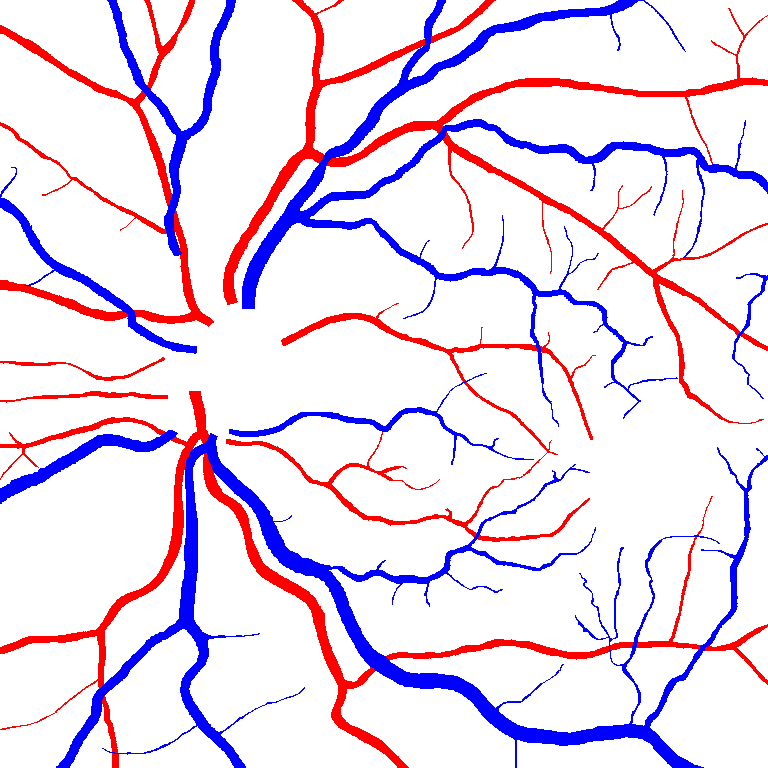}
\hfill
\includegraphics[width=\x\linewidth,height=\x\linewidth]{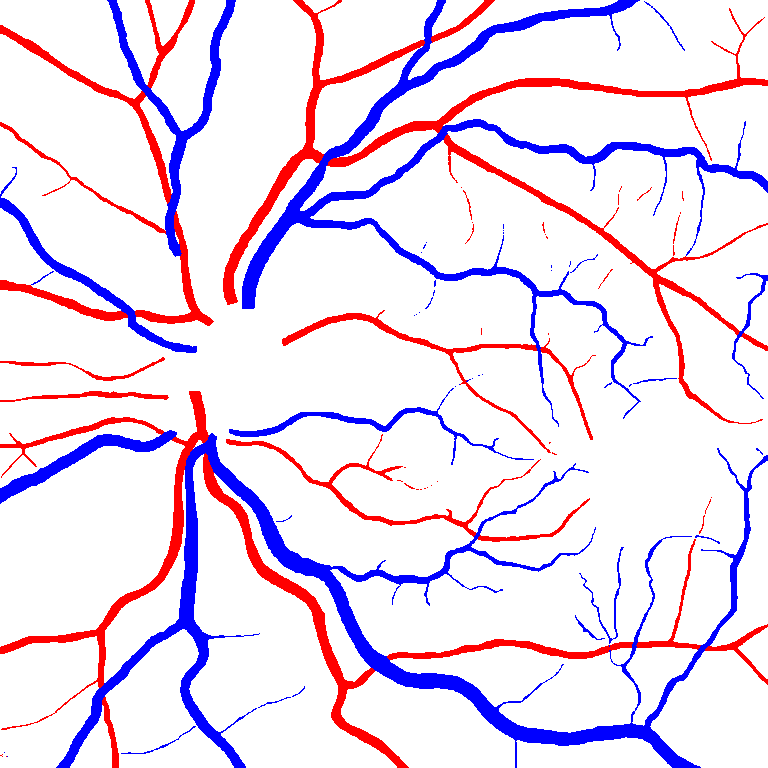}
\hfill
\includegraphics[width=\x\linewidth,height=\x\linewidth]{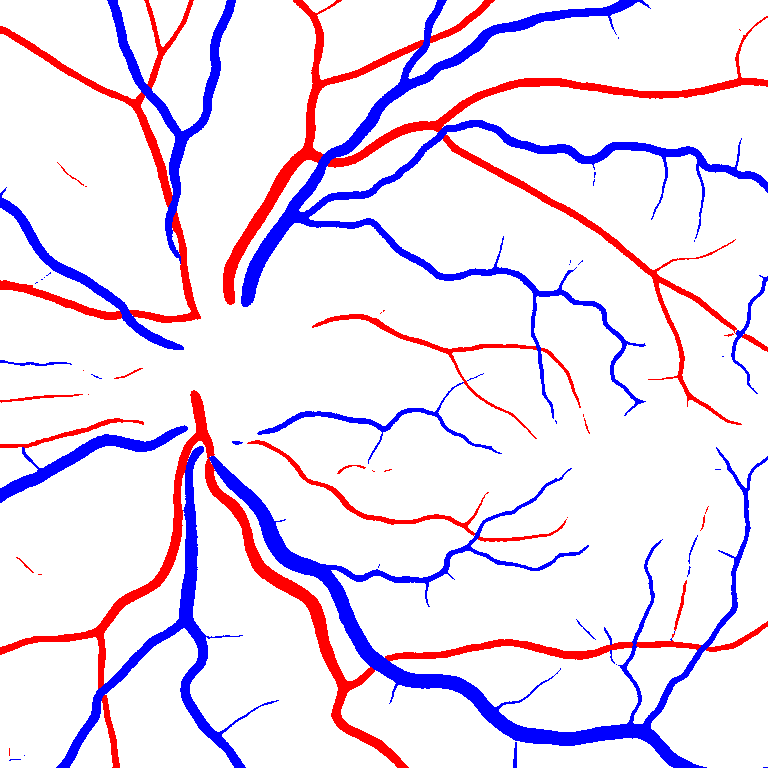}
\hfill
\includegraphics[width=\x\linewidth,height=\x\linewidth]{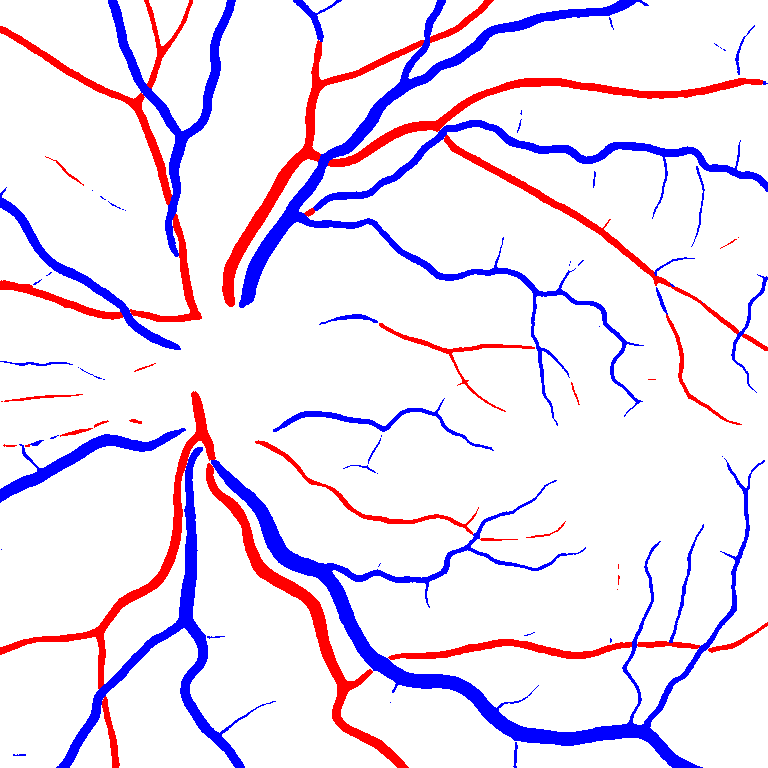}
\hfill
\includegraphics[width=\x\linewidth,height=\x\linewidth]{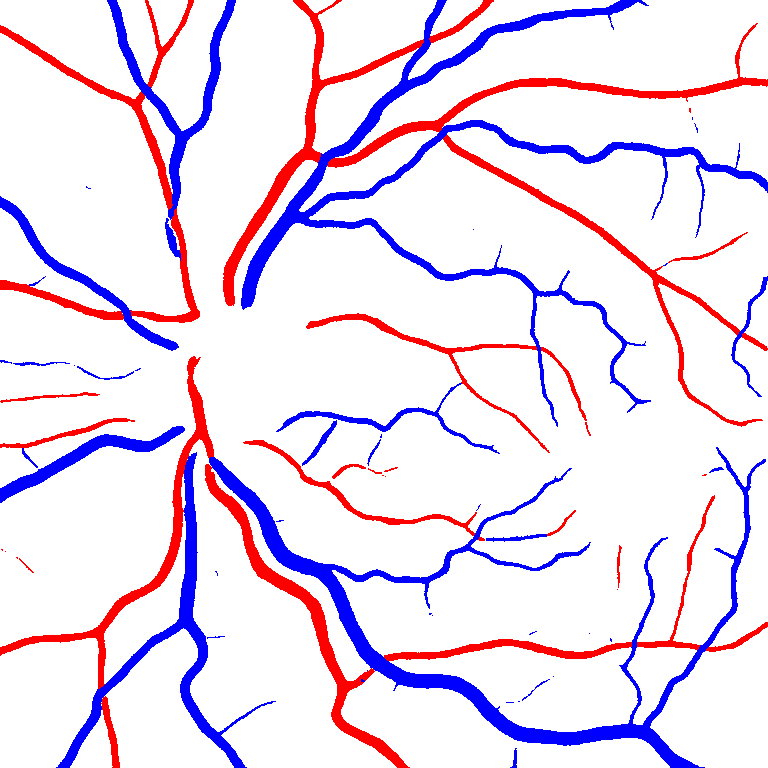}
\\[10pt]
\includegraphics[width=\x\linewidth,height=\x\linewidth]{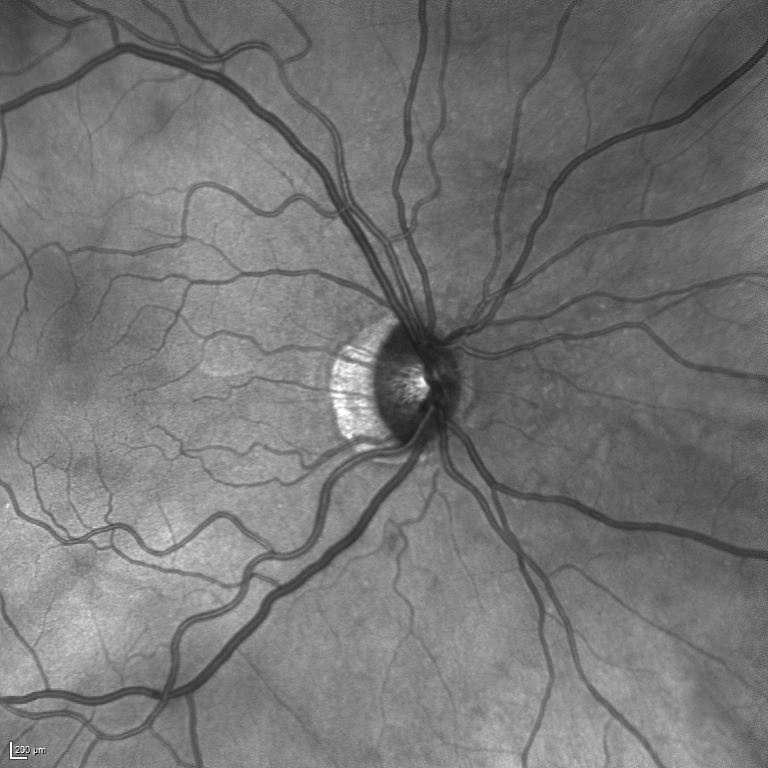}
\hfill
\includegraphics[width=\x\linewidth,height=\x\linewidth]{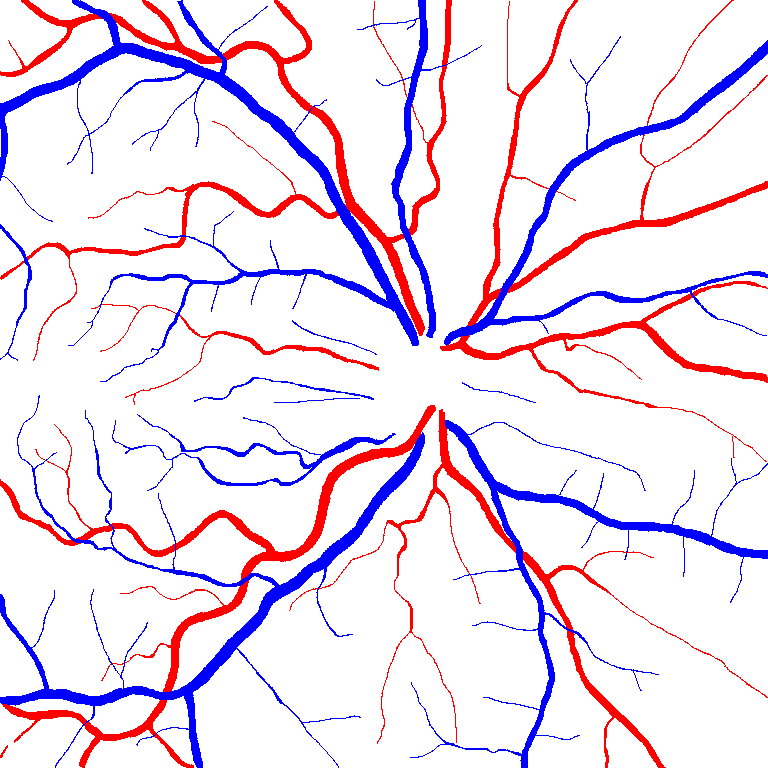}
\hfill
\includegraphics[width=\x\linewidth,height=\x\linewidth]{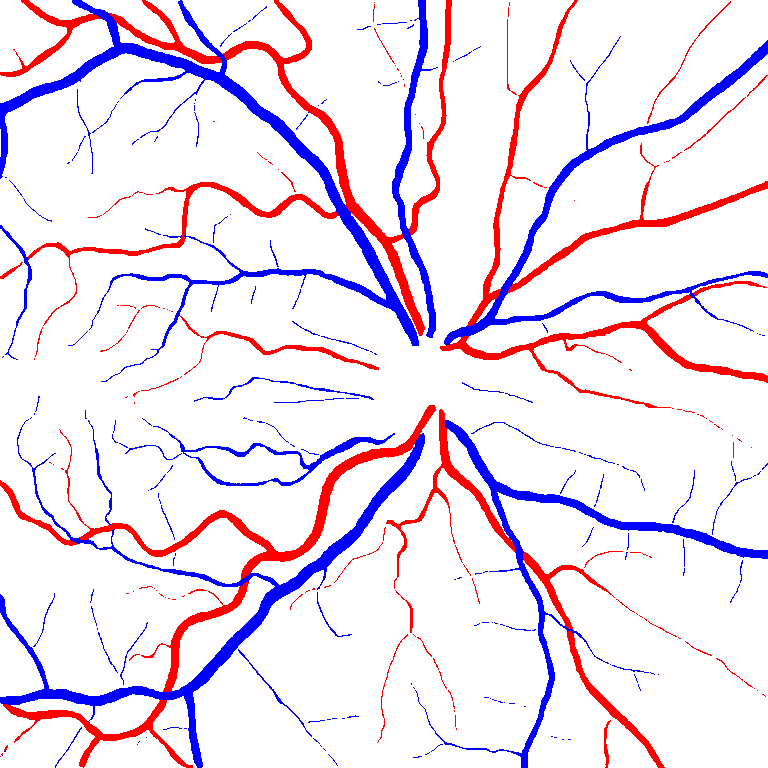}
\hfill
\includegraphics[width=\x\linewidth,height=\x\linewidth]{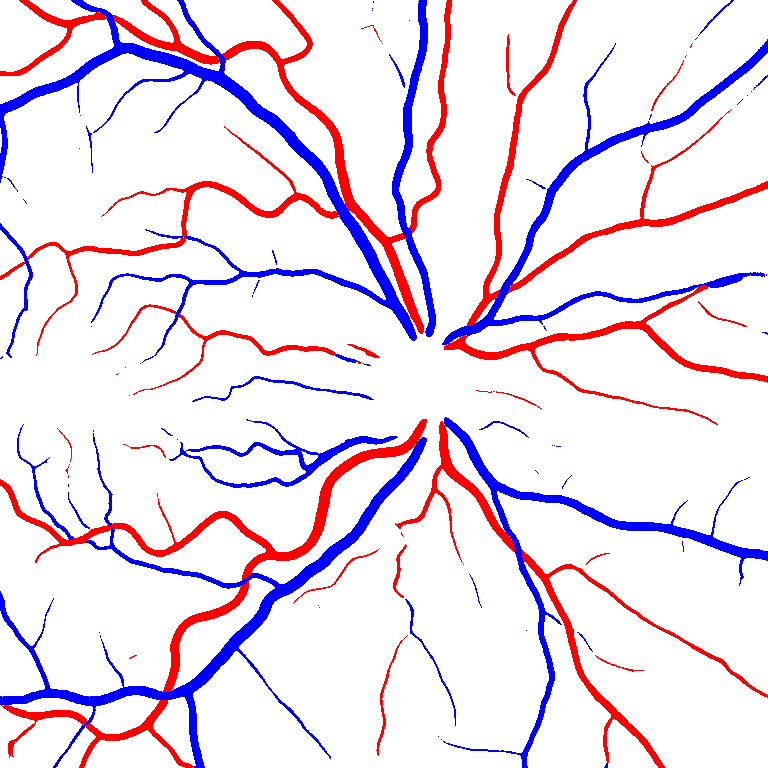}
\hfill
\includegraphics[width=\x\linewidth,height=\x\linewidth]{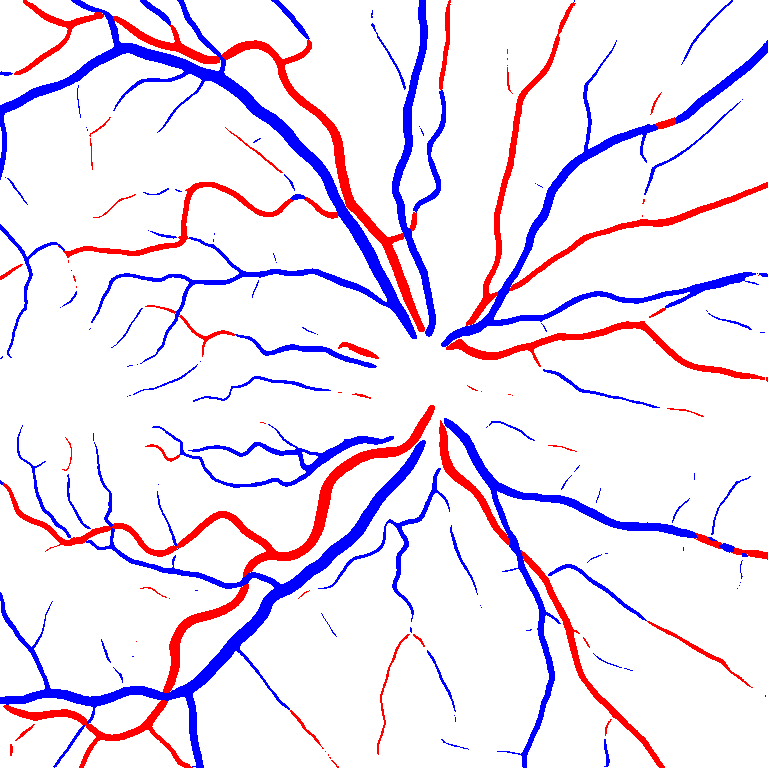}
\hfill
\includegraphics[width=\x\linewidth,height=\x\linewidth]{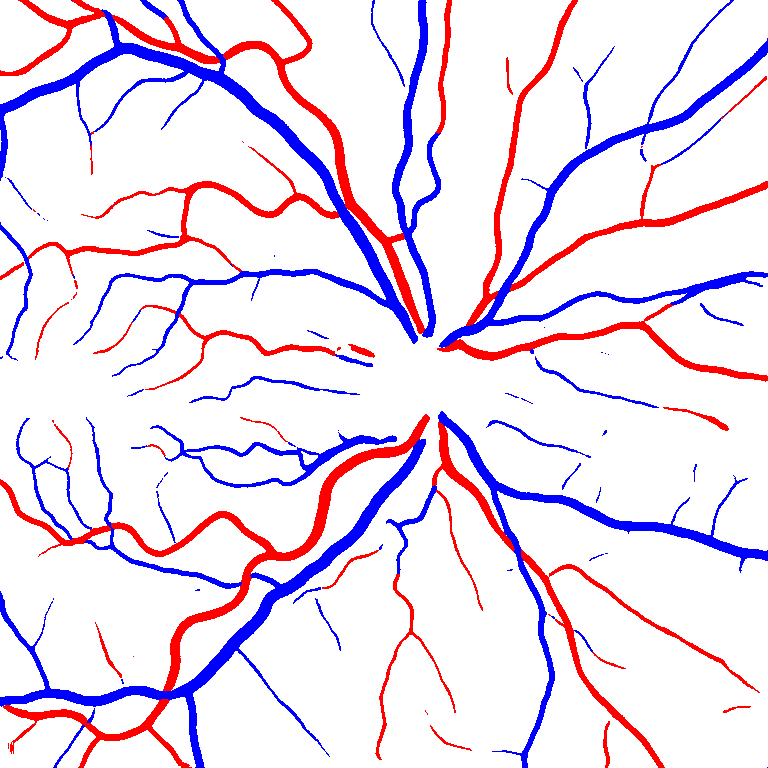}
\\[10pt]
\includegraphics[width=\x\linewidth,height=\x\linewidth]{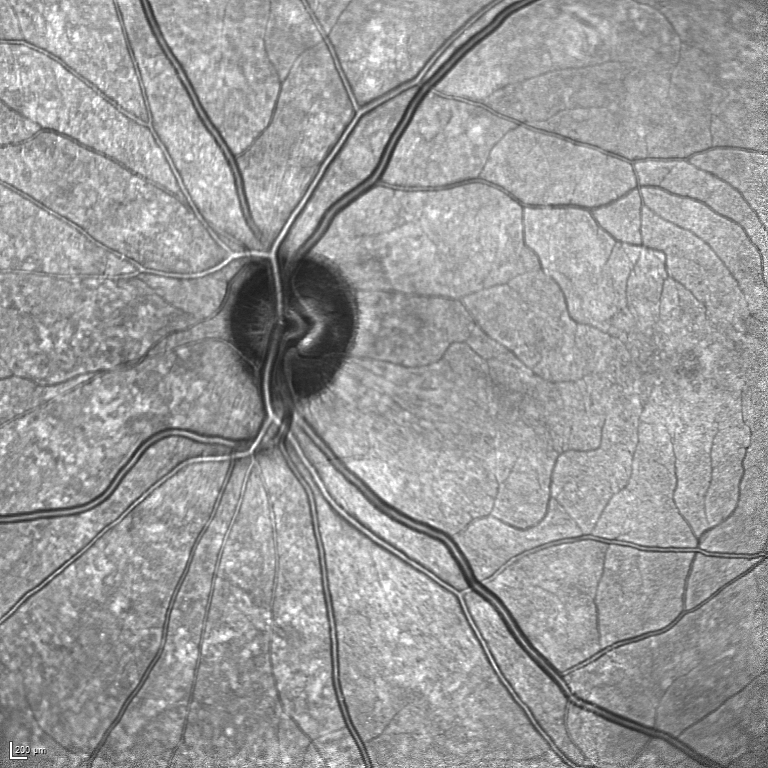}
\hfill
\includegraphics[width=\x\linewidth,height=\x\linewidth]{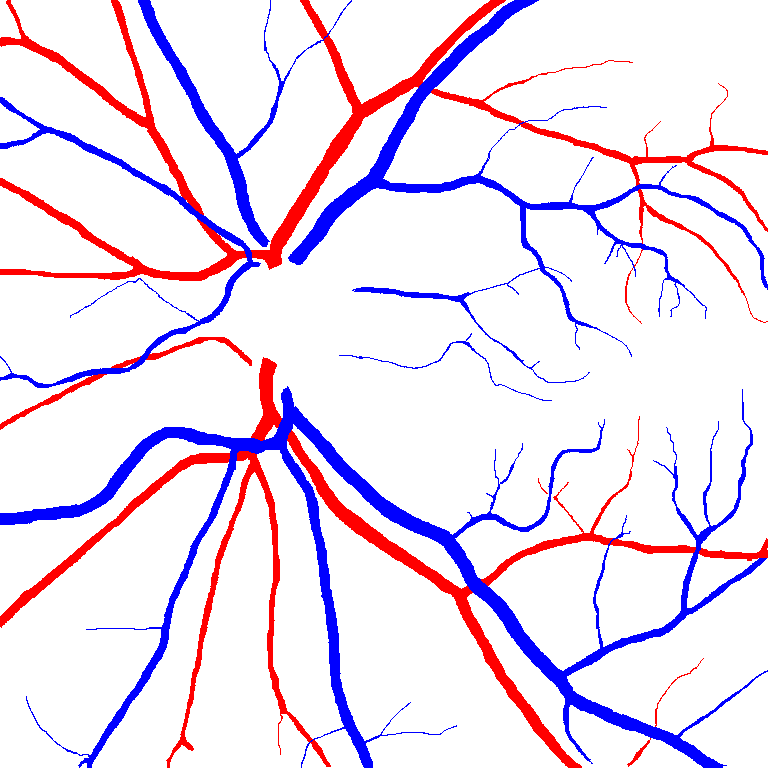}
\hfill
\includegraphics[width=\x\linewidth,height=\x\linewidth]{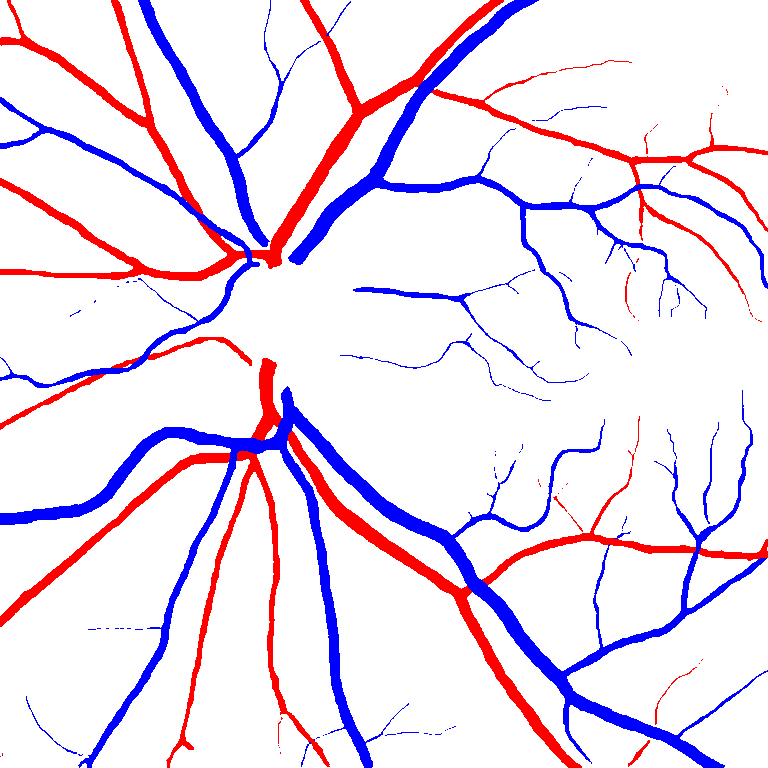}
\hfill
\includegraphics[width=\x\linewidth,height=\x\linewidth]{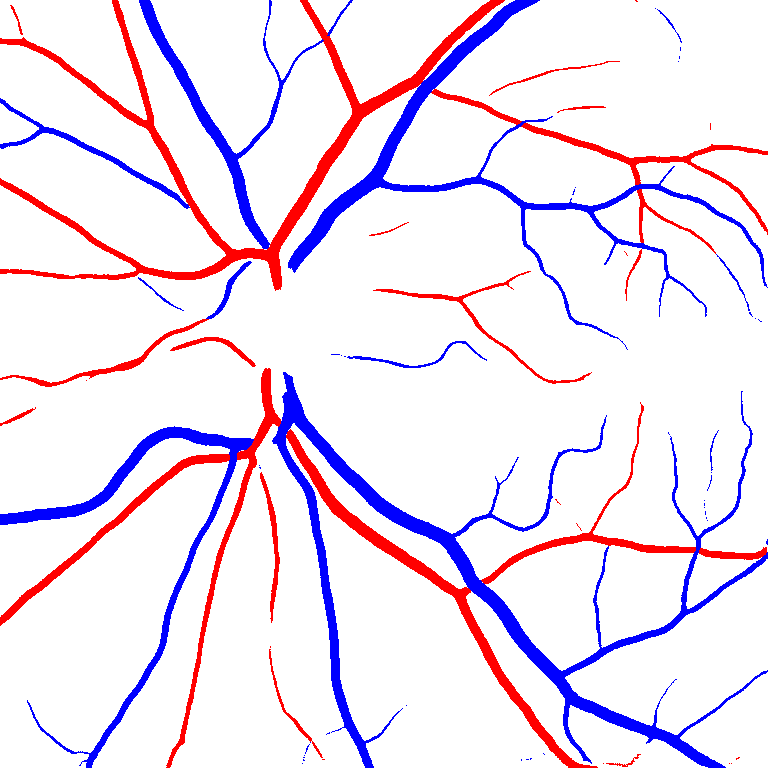}
\hfill
\includegraphics[width=\x\linewidth,height=\x\linewidth]{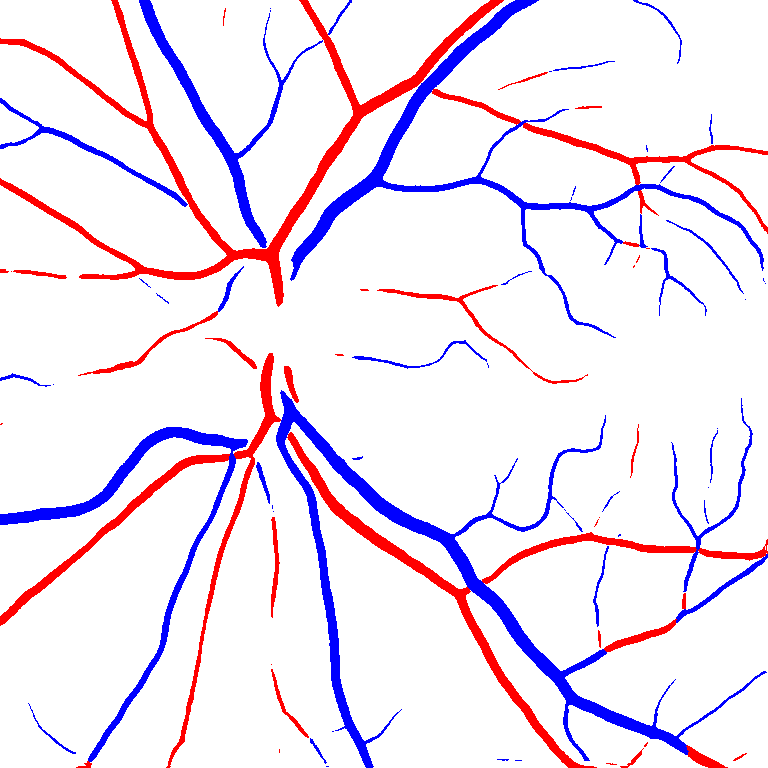}
\hfill
\includegraphics[width=\x\linewidth,height=\x\linewidth]{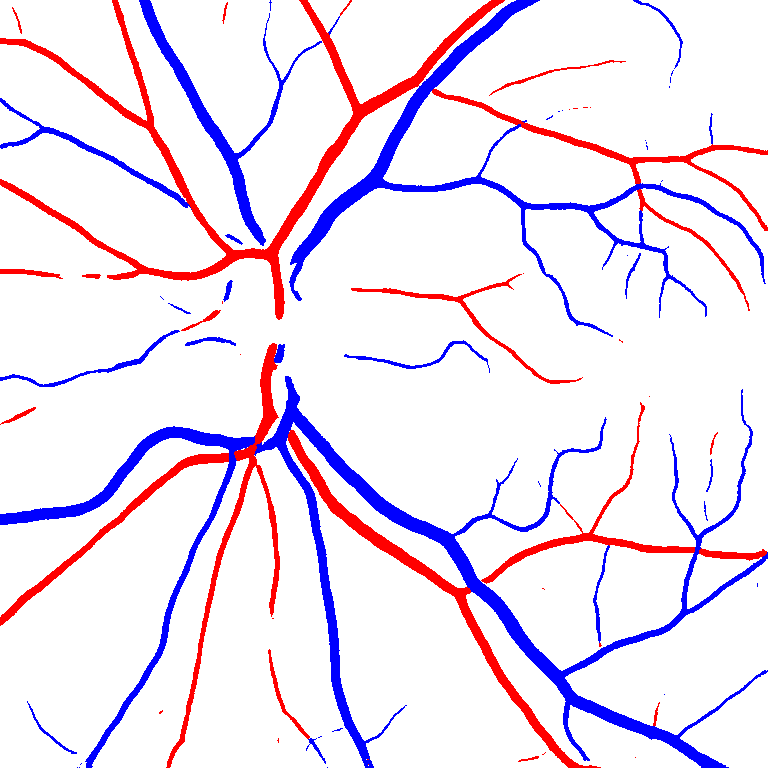}
\\[10pt]
\includegraphics[width=\x\linewidth,height=\x\linewidth]{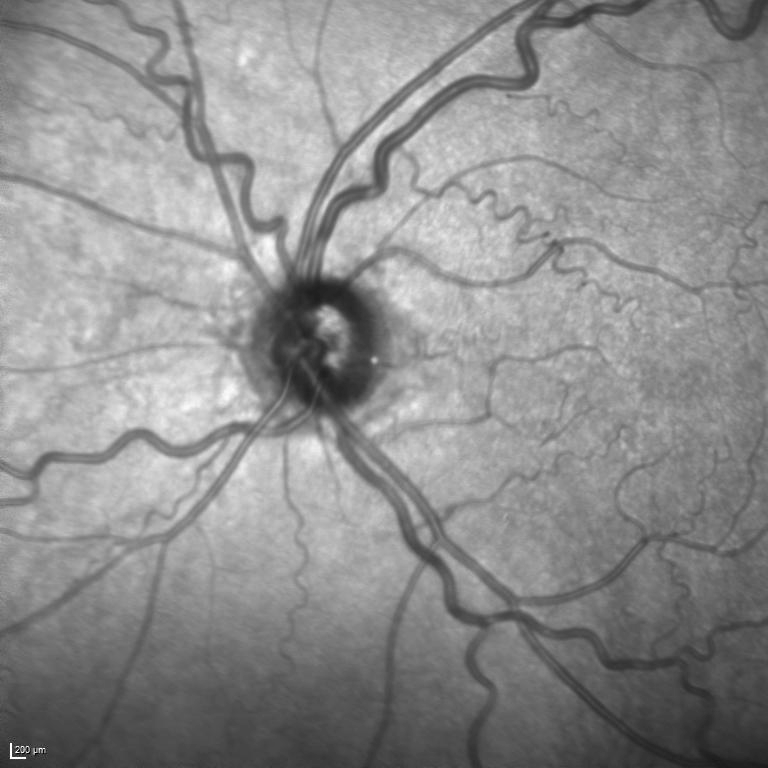}
\hfill
\includegraphics[width=\x\linewidth,height=\x\linewidth]{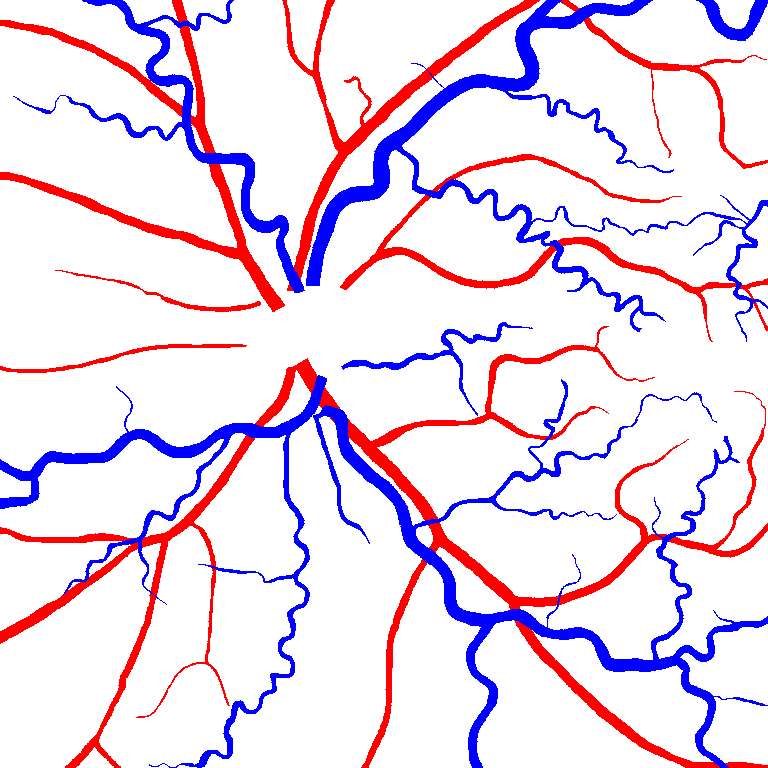}
\hfill
\includegraphics[width=\x\linewidth,height=\x\linewidth]{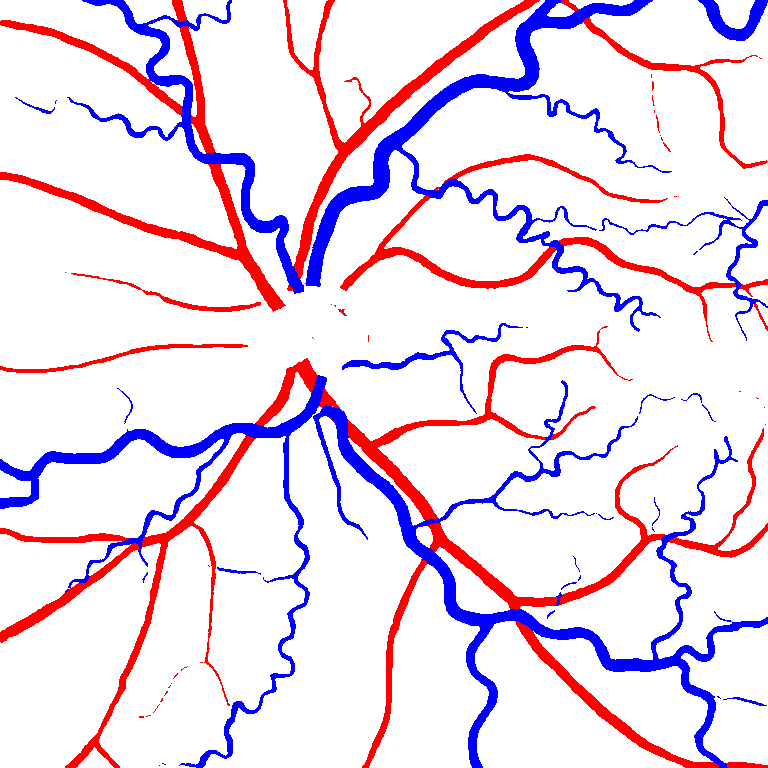}
\hfill
\includegraphics[width=\x\linewidth,height=\x\linewidth]{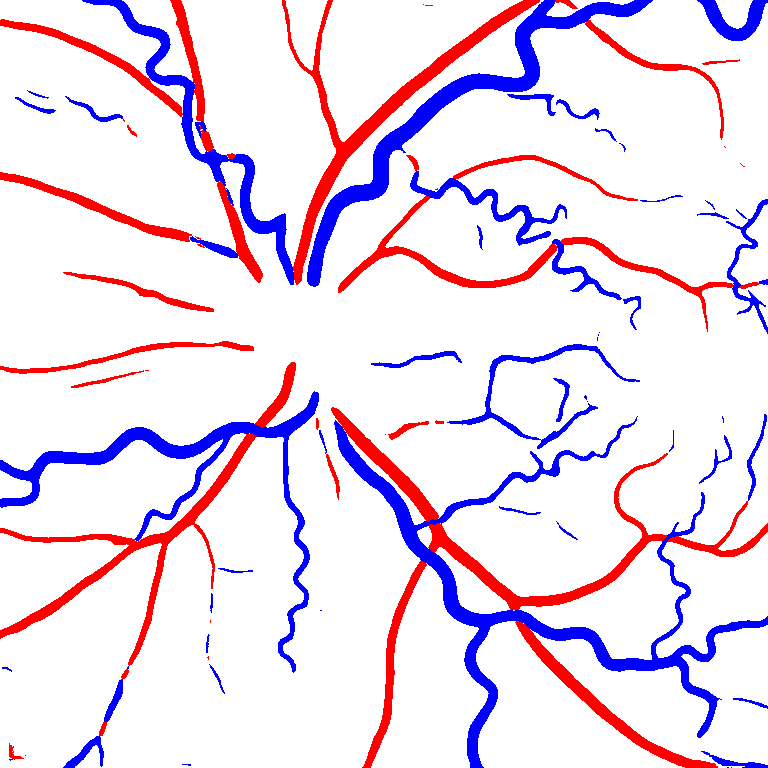}
\hfill
\includegraphics[width=\x\linewidth,height=\x\linewidth]{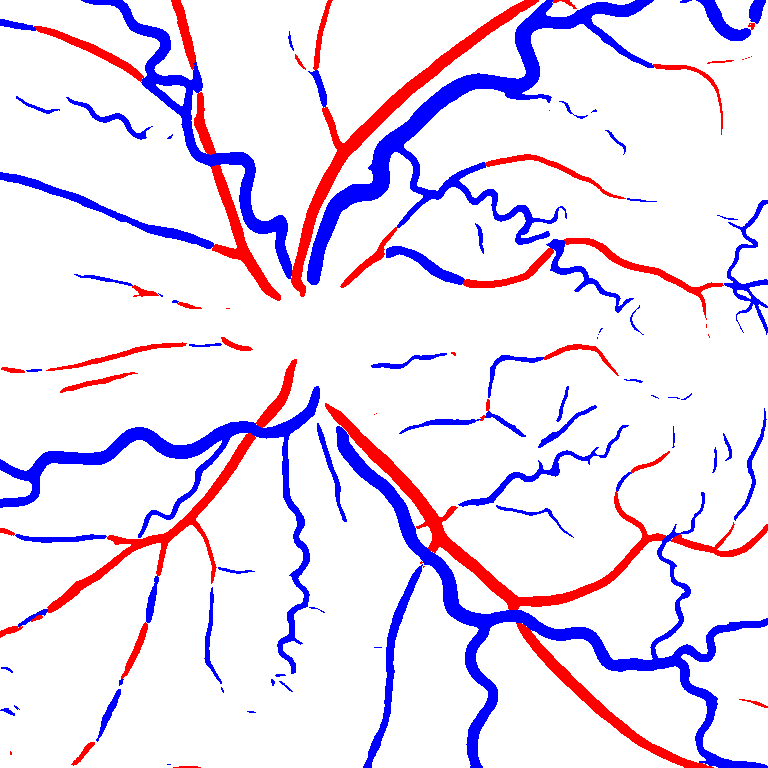}
\hfill
\includegraphics[width=\x\linewidth,height=\x\linewidth]{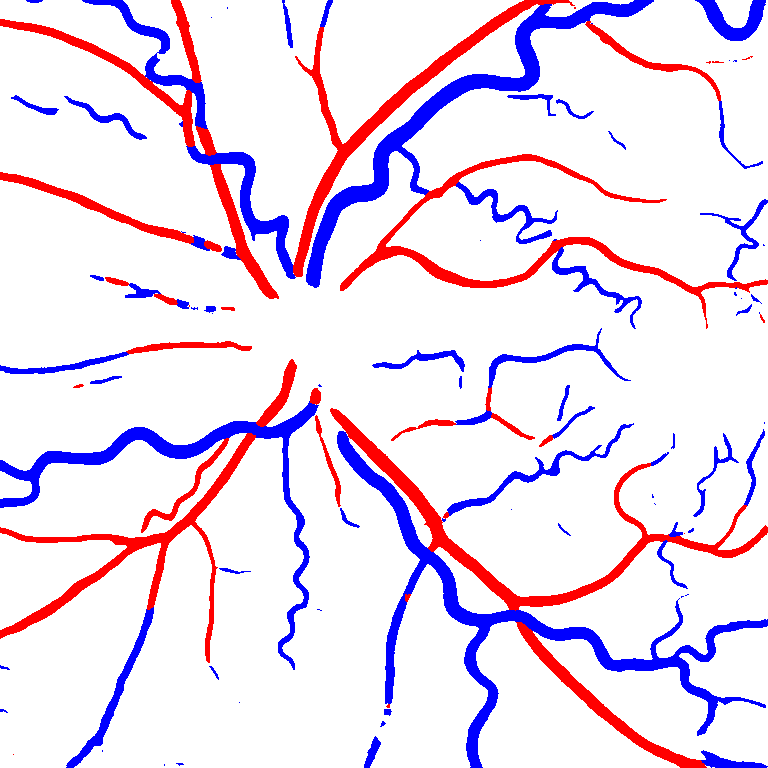}
\\[10pt]
\makebox[\x\linewidth]{(a)} \hfill \makebox[\x\linewidth]{(b)} \hfill
\makebox[\x\linewidth]{(c)} \hfill \makebox[\x\linewidth]{(d)} \hfill \makebox[\x\linewidth]{(e)} \hfill \makebox[\x\linewidth]{(f)} \hfill
\caption{Qualitative comparison of segmentation outputs on the RAVIR test set. (a) Input image. (b) Semantic Labels. (c) SegRAVIR. (d) CE-Net. (e) R2U-Net. (f) U-Net. The artery and vein classes are indicated by red and blue, respectively.}
\label{fig:qualseg}
\end{figure*}

\begin{table*} \centering
\caption{Quantitative comparison of semantic segmentation performance on the RAVIR dataset}
\label{tab_res_avir}
\begin{tabular}{lccccc ccccc c}
\toprule
 & \multicolumn{5}{c}{Artery} & \multicolumn{5}{c}{Vein}& \multicolumn{1}{c}{Average}\\
\cmidrule(lr){2-6} \cmidrule(lr){7-11} 
Method & SE & SP & Acc & AUC&Dice & SE & SP & Acc& AUC&Dice&Dice\\
\midrule
U-Net~\cite{ronneberger2015u} & 0.7275 & 0.9788 & 0.9670 & 0.9665 & 0.7510 & 0.7389 & 0.9795 & 0.9685&0.9713&0.7674&0.7592\\
Dense U-Net~\cite{guan2019fully} & 0.7296 & 0.9795 &  0.9681 & 0.9673 &  0.7584 &0.7410 & 0.9808 &  0.9691 & 0.9736 &  0.7704 &  0.7644\\
Residual U-Net~\cite{alom2019recurrent} & 0.7375 & 0.9828 &  0.9697 & 0.9689 &  0.7602 
&0.7455 & 0.9839 &  0.9701 & 0.9773 &  0.7768 &  0.7685\\
R2U-Net~\cite{alom2019recurrent} & 0.7408 & 0.9810 & 0.9706 & 0.9727 & 0.7621 & 0.7492 & 0.9842 & 0.9728&0.9802&0.7826&0.7723\\
Recurrent U-Net~\cite{alom2019recurrent}  & 0.7389 & 0.9833 &  0.9718 & 0.9749 &  0.7690&0.7478 & 0.9849 &  0.9735 & 0.9810 &  0.7898 &  0.7794\\
U-Net++~\cite{zhou2019unetplusplus} & 0.7406 & 0.9827 &  0.9729 & 0.9772 &  0.7781&0.7527 & 0.9837 &  0.9743 & 0.9821 &  0.7909 &  0.7845\\
DU-Net~\cite{jin2019dunet} & 0.7432 & 0.9854 &  0.9750 & 0.9785 &  0.7833&0.7569 & 0.9870 &  0.9762 & 0.9840 &  0.7972 &  0.7902\\
AG-Net~\cite{zhang2019attention} & 0.7426 & 0.9860 &  0.9751 & 0.9792 &  0.7854&0.7562 & 0.9879 &  0.9768 & 0.9844 &  0.7983 &  0.7918\\
IterNet~\cite{Li_2020_WACV} & 0.7431 & 0.9863 & 0.9765 & 0.9820 & 0.7898 & 0.7586 & 0.9895 & 0.9779&0.9852&0.8009&0.7953\\
CE-Net~\cite{gu2019net} & 0.7501 & 0.9881 & 0.9771 & 0.9834 & 0.7928 & 0.7682 & 0.9908 & 0.9786&0.9871&0.8053&0.7990\\
\textbf{SegRAVIR} &\textbf{0.7772} & \textbf{0.9925} &  \textbf{0.9815} & \textbf{0.9902} &  \textbf{0.8287}&\textbf{0.8086} & \textbf{0.9951} &  \textbf{0.9829} & \textbf{0.9916} &  \textbf{0.8301} &  \textbf{0.8294}\\
\bottomrule
\end{tabular}
\end{table*}

\begin{table} \centering
\caption{Quantitative comparison of semantic segmentation performance on the RITE dataset}
\label{tab:rite}
\begin{tabular}{llcc}
\toprule
Method  & SE & SP & Acc\\
\midrule
Dashtbozorg \textit{et al.}~\cite{dashtbozorg2013automatic}& 0.9000&0.8400&	0.8740	 \\
Estrada \textit{et al.}~\cite{estrada2015retinal}& 0.9300&0.9410&	0.9350	 \\
Girard \textit{et al.}~\cite{girard2019joint}& 0.8630&0.8660&	0.8650	 \\
Hemelings \textit{et al.}~\cite{hemelings2019artery}& -&-&	0.9487	  \\
Ma \textit{et al.}~\cite{ma2019multi}& 0.9220&0.9300&	0.9260	  \\
Kang \textit{et al.}~\cite{kang2020avnet}& 0.8863&0.9272&	0.9081	  \\
Morano \textit{et al.}~\cite{morano2021simultaneous}& 0.8747&0.9089&0.8924	  \\\textbf{SegRAVIR}& \textbf{0.9310}&	\textbf{0.9431}&\textbf{0.9513}	  \\
\bottomrule
\end{tabular}
\end{table}

\subsection{Evaluation Metrics}

\subsubsection{Segmentation}

\def\RE{\textit{RE}}
\def\PR{\textit{PR}}
\def\SP{\textit{SP}}
\def\SE{\textit{SE}}
\def\TP{\textit{TP}}
\def\TN{\textit{TN}}
\def\FP{\textit{FP}}
\def\FN{\textit{FN}}
\def\AUC{\textit{AUC}}
\def\F1{\textit{F1}}
\def\Acc{\textit{Acc}}

To assess the accuracy of model predictions, we used various evaluation metrics, including Sensitivity (\SE), Specificity (\SP), Accuracy (\Acc), Precision (\PR), Recall (\RE), Area Under the Receiver Operating Characteristics (ROC)  Curve (\AUC), and the \F1 score, which is the same as the Dice coefficient in the context of image segmentation. 

\begin{table} \centering
\caption{Measured diameters (in microns) of segmented arteries and veins in the RAVIR test set}
\label{tab_res_width}
\begin{tabular}{l cc cc}
\toprule
 & \multicolumn{2}{c}{Artery} & \multicolumn{2}{c}{Vein}\\
\cmidrule(lr){2-3} \cmidrule(lr){4-5} 
Method & Measured & MAPE & Measured & MAPE\\
\midrule
Dense U-Net~\cite{guan2019fully} &  99.98$\pm$8.12&0.0618&112.54$\pm$9.89&0.0607\\
U-Net~\cite{ronneberger2015u} &99.93$\pm$7.31&0.0612&112.42$\pm$9.76&0.0595\\
Residual U-Net~\cite{alom2019recurrent}&99.88$\pm$6.18&0.0596&112.30$\pm$10.14&0.0586\\
Recurrent U-Net~\cite{alom2019recurrent}&99.76$\pm$6.53&0.0583&112.57$\pm$8.59&0.0535\\
R2U-Net~\cite{alom2019recurrent}&96.94$\pm$6.14&0.0578&113.86$\pm$9.85&0.0526\\
U-Net++~\cite{zhou2019unetplusplus} &   93.96$\pm$5.64&0.0559&111.63$\pm$10.83&0.0520\\
AG-Net~\cite{zhang2019attention} &   94.34$\pm$6.76&0.0548&110.84$\pm$9.42&0.0514\\
DU-Net~\cite{jin2019dunet}&97.81$\pm$5.54&0.0539&110.57$\pm$9.48&0.0501\\
IterNet~\cite{Li_2020_WACV} &  97.54$\pm$6.10&0.0502&111.55$\pm$9.77&0.0498\\
CE-Net~\cite{gu2019net} &  97.33$\pm$5.75&0.0463&113.73$\pm$11.23&0.0457\\
\textbf{SegRAVIR} & \textbf{98.29$\pm$5.48}&\textbf{0.0409}&\textbf{112.34$\pm$10.02}&\textbf{0.0393}\\
\bottomrule\\[-2pt]
\multicolumn{5}{l}{MAPE: Mean Absolute Percentage Error}
\end{tabular}
\end{table}

\subsubsection{Width Estimation}

We refer to the diameter of the vessel as twice the maximum distance to its medial axis. To asses the accuracy of diameter estimations, we employ the Mean Absolute Percentage Error (MAPE) defined as
\begin{equation}
\textit{MAPE}=\dfrac{1}{N}\sum_{m=1}^{N}\left |\frac{y_{m}-\hat y_{m}}{y_{m}}\right |
\label{eq:MAPE},
\end{equation}
where $N$, $\hat y_{m}$, and $y_{m}$ denote the total number of cases, predictions, and ground truth values, respectively.

\section{Results}

In this section, we first present quantitative and qualitative comparisons between the SegRAVIR model and competing methodologies on the RAVIR dataset as well as on the RITE dataset~\cite{hu2013automated}. Then we report the pixel-wise diameter profiles of SegRAVIR-segmented arteries and veins with quantitative error analysis. Finally, we present the benchmarks of pretrained RAVIR networks on three publicly available color datasets, the DRIVE~\cite{staal2004ridge}, STARE~\cite{hoover2000locating}, and CHASE\_DB1~\cite{fraz2012ensemble} datasets, and demonstrate the effectiveness of our proposed distillation methodology.

\subsection{Retinal Artery and Vein Segmentation}
\label{sec:res1}

As reported in Table~\ref{tab_res_avir}, we compared the SegRAVIR model against competing deep learning-based segmentation approaches on the RAVIR dataset. Evidently, SegRAVIR outperforms these methods as judged by all metrics for artery and vein classes with a healthy margin. In terms of Dice score, SegRAVIR outperforms CE-NET, IterNet and AG-Net by $16.28\%$, $26.71\%$ and $27.48\%$ for artery segmentation and by $16.28\%$, $26.71\%$ and $27.48\%$ for vein segmentation, respectively. Fig.~\ref{fig:qualseg} presents a qualitative comparison of the semantic segmentation outputs of SegRAVIR, CE-Net, and U-Net. Specifically, SegRAVIR yields more accurate vessel topology (i.e., thickness and orientation) segmentation with higher pixel-wise classification accuracy.

\begin{figure}
\centering
{\includegraphics[width=0.32\linewidth,height=0.32\linewidth]{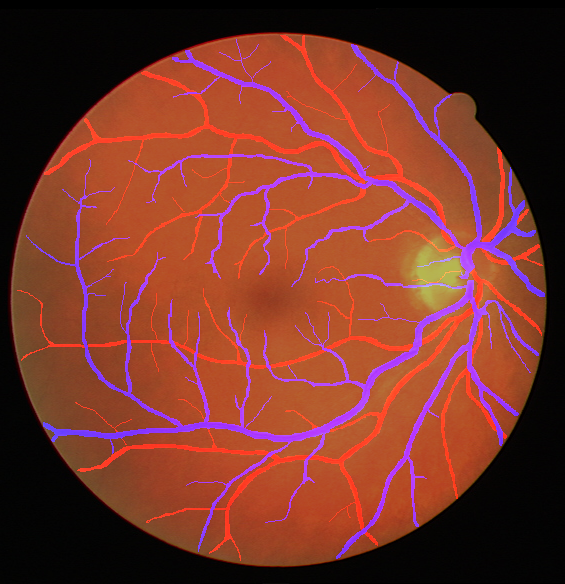}}
\hfill
{\includegraphics[width=0.32\linewidth,height=0.32\linewidth]{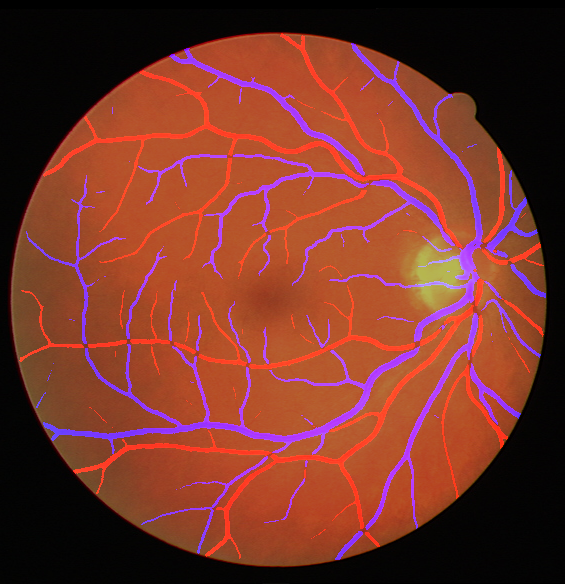}}
\hfill
{\includegraphics[width=0.32\linewidth,height=0.32\linewidth]{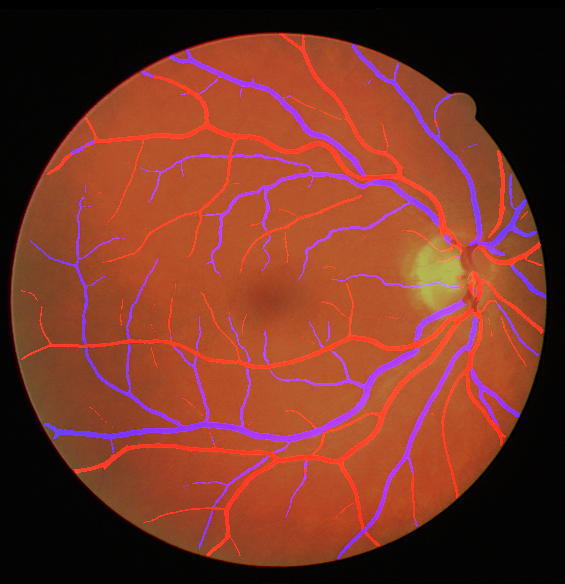}}
\hfill\\[6pt]
{\includegraphics[width=0.32\linewidth,height=0.32\linewidth]{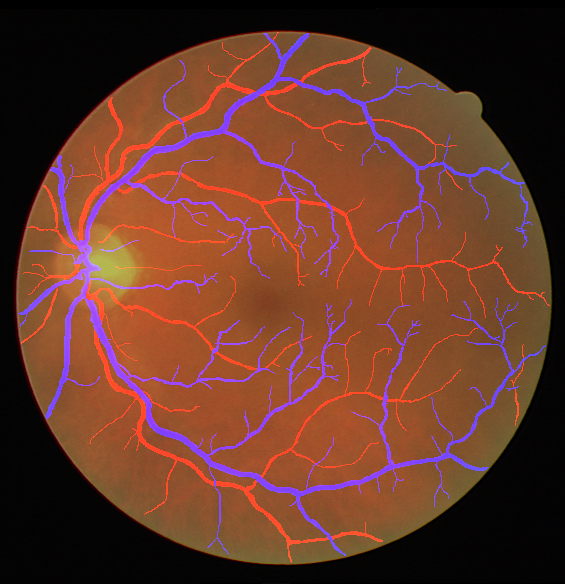}}
\hfill
{\includegraphics[width=0.32\linewidth,height=0.32\linewidth]{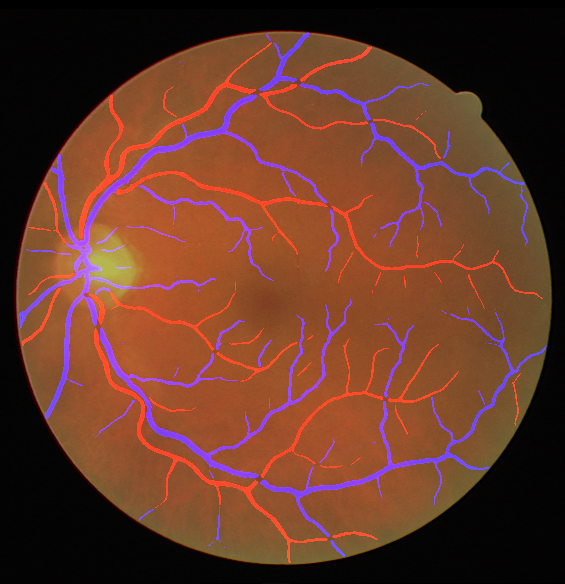}}
\hfill
{\includegraphics[width=0.32\linewidth,height=0.32\linewidth]{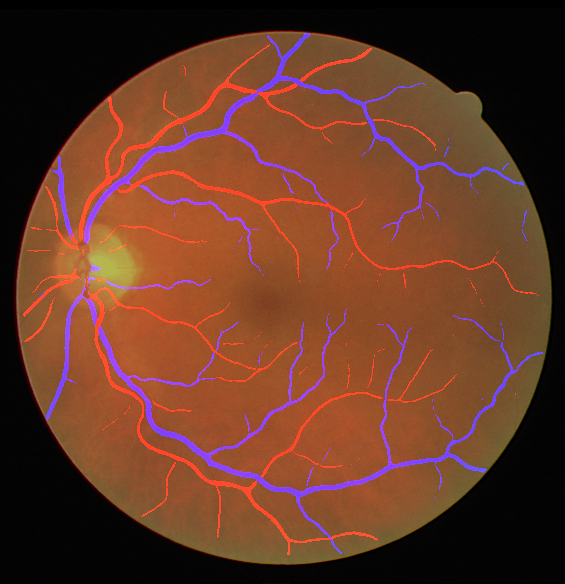}}
\hfill \\[6pt]
\makebox[0.32\linewidth]{(a)}
\makebox[0.32\linewidth]{(b)} 
\makebox[0.32\linewidth]{(c)}
\caption{Qualitative comparison of segmentation outputs on the RITE test set. (a) Reference labels. (b) SegRAVIR. (c) Hemelings \textit{et al.}~\cite{hemelings2019artery}. The artery and vein classes are indicated by red and blue, respectively.}
\label{fig:rite_qual}
\end{figure}

Table~\ref{tab:rite} presents quantitative performance benchmarks of SegRAVIR and other competing approaches for retinal artery and vein classification on the RITE dataset~\cite{hu2013automated}. SegRAVIR outperforms previous state-of-the-art approaches in terms of accuracy, sensitivity, and specificity. Fig.~\ref{fig:rite_qual} provides a qualitative comparison between segmentation outputs of SegRAVIR and the method of Hemelings~\textit{et al.}~\cite{hemelings2019artery} on the RITE test set.

\subsection{Vessel Width Estimation}

\begin{figure}
\centering
{\includegraphics[width=0.32\linewidth,height=0.32\linewidth]{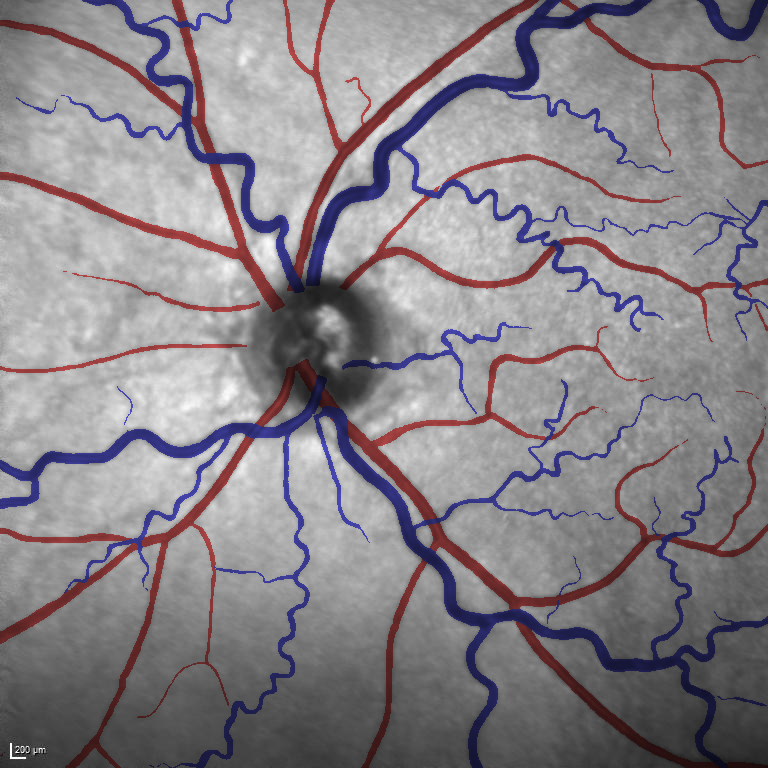}}
\hfill
{\includegraphics[width=0.32\linewidth,height=0.32\linewidth]{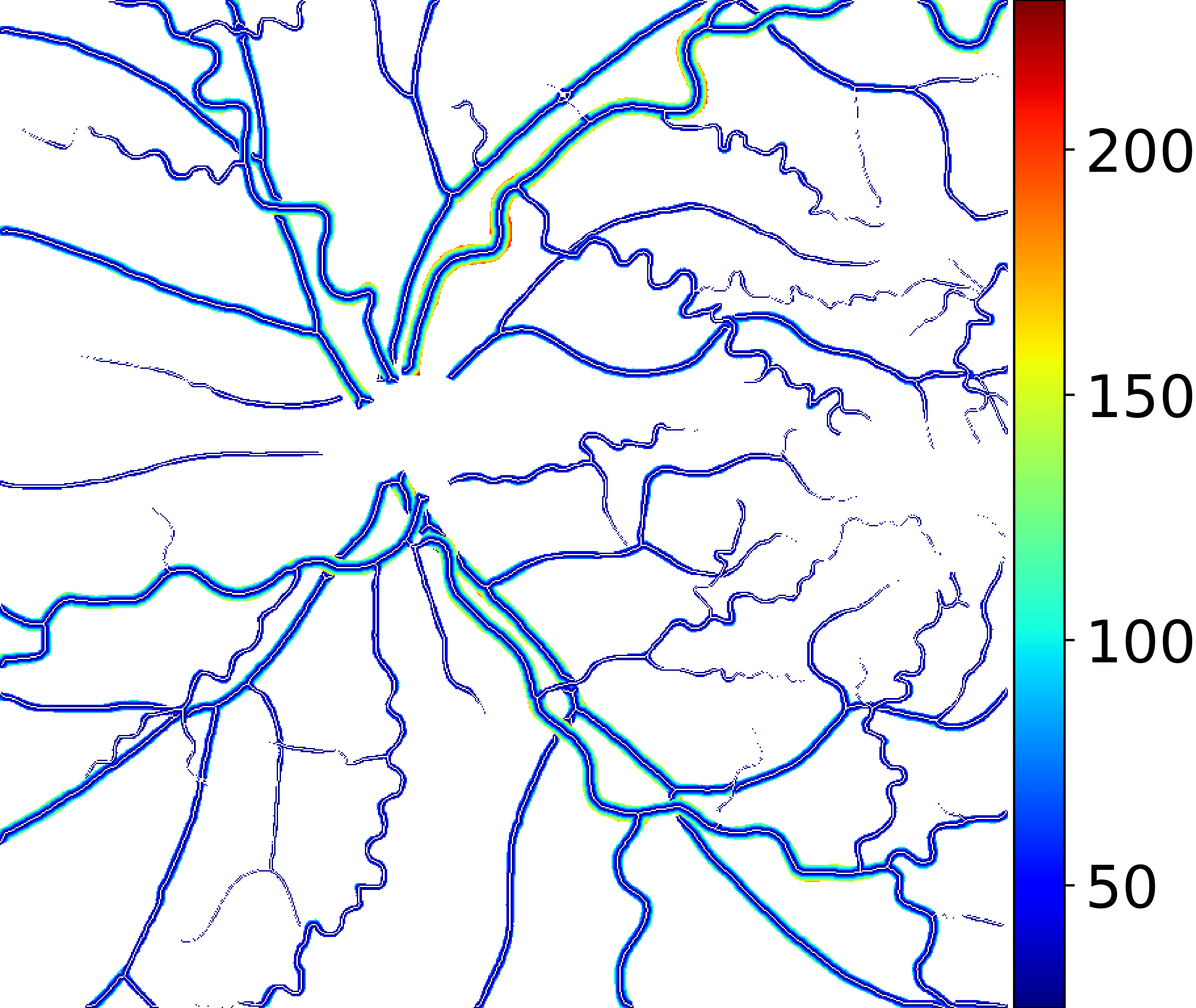}}
\hfill
{\includegraphics[width=0.32\linewidth,height=0.32\linewidth]{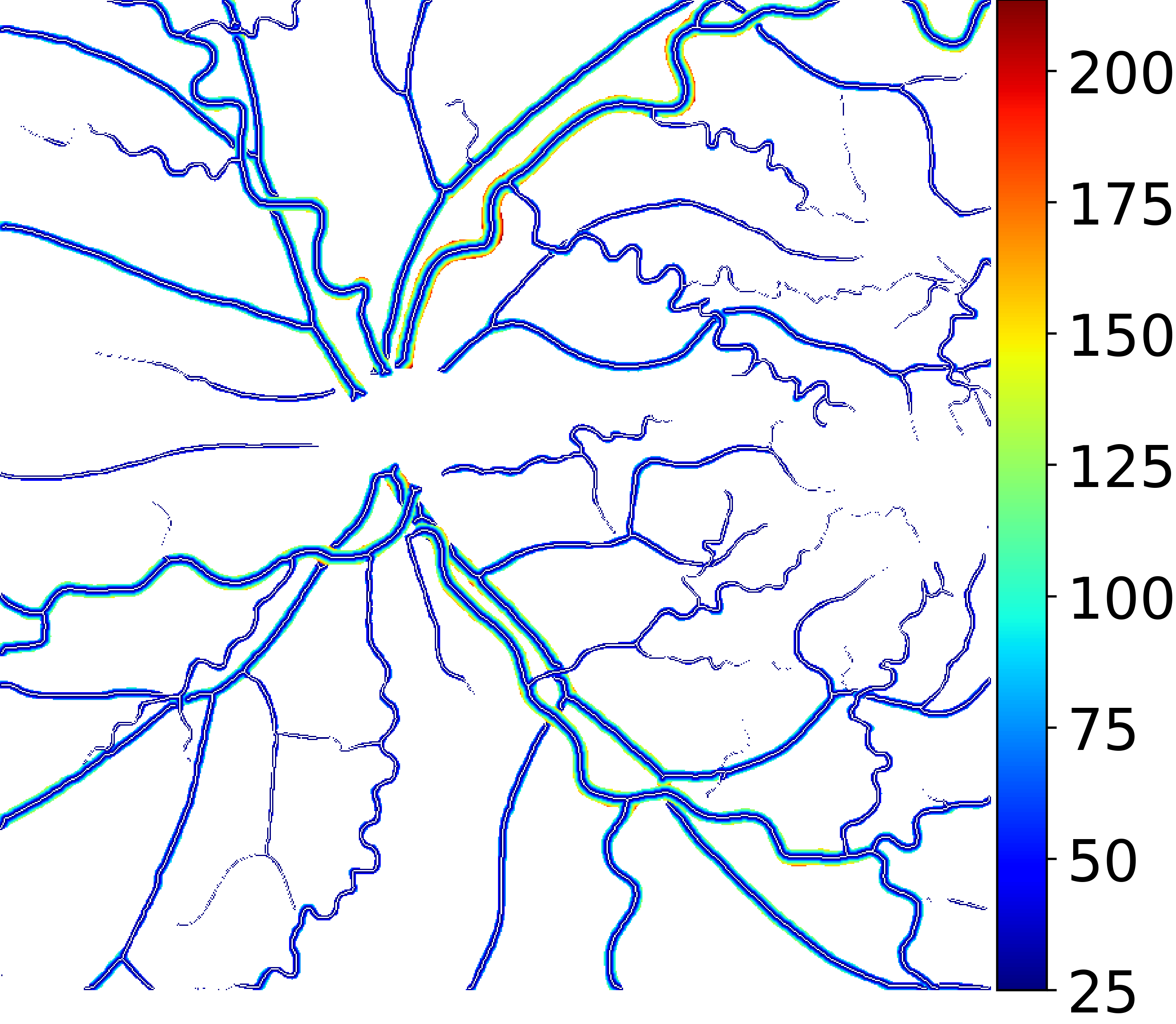}}
\hfill \\[6pt]
{\includegraphics[width=0.32\linewidth,height=0.32\linewidth]{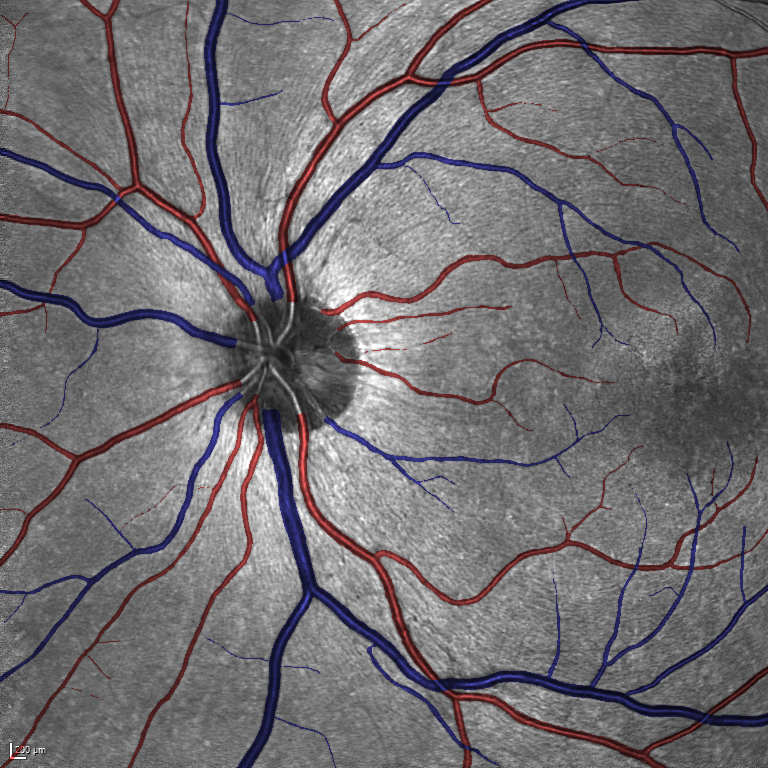}}
\hfill
{\includegraphics[width=0.32\linewidth,height=0.32\linewidth]{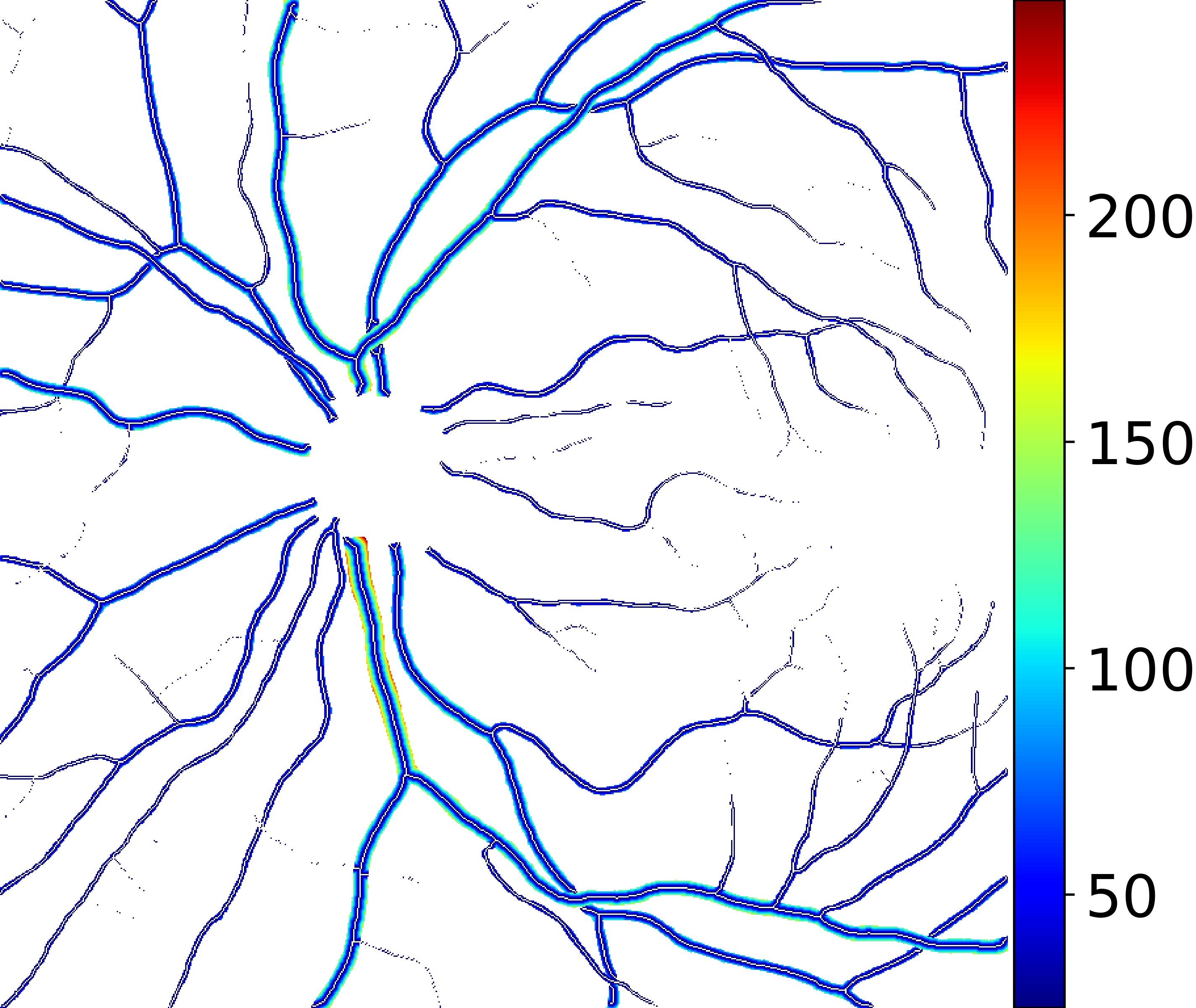}}
\hfill
{\includegraphics[width=0.32\linewidth,height=0.32\linewidth]{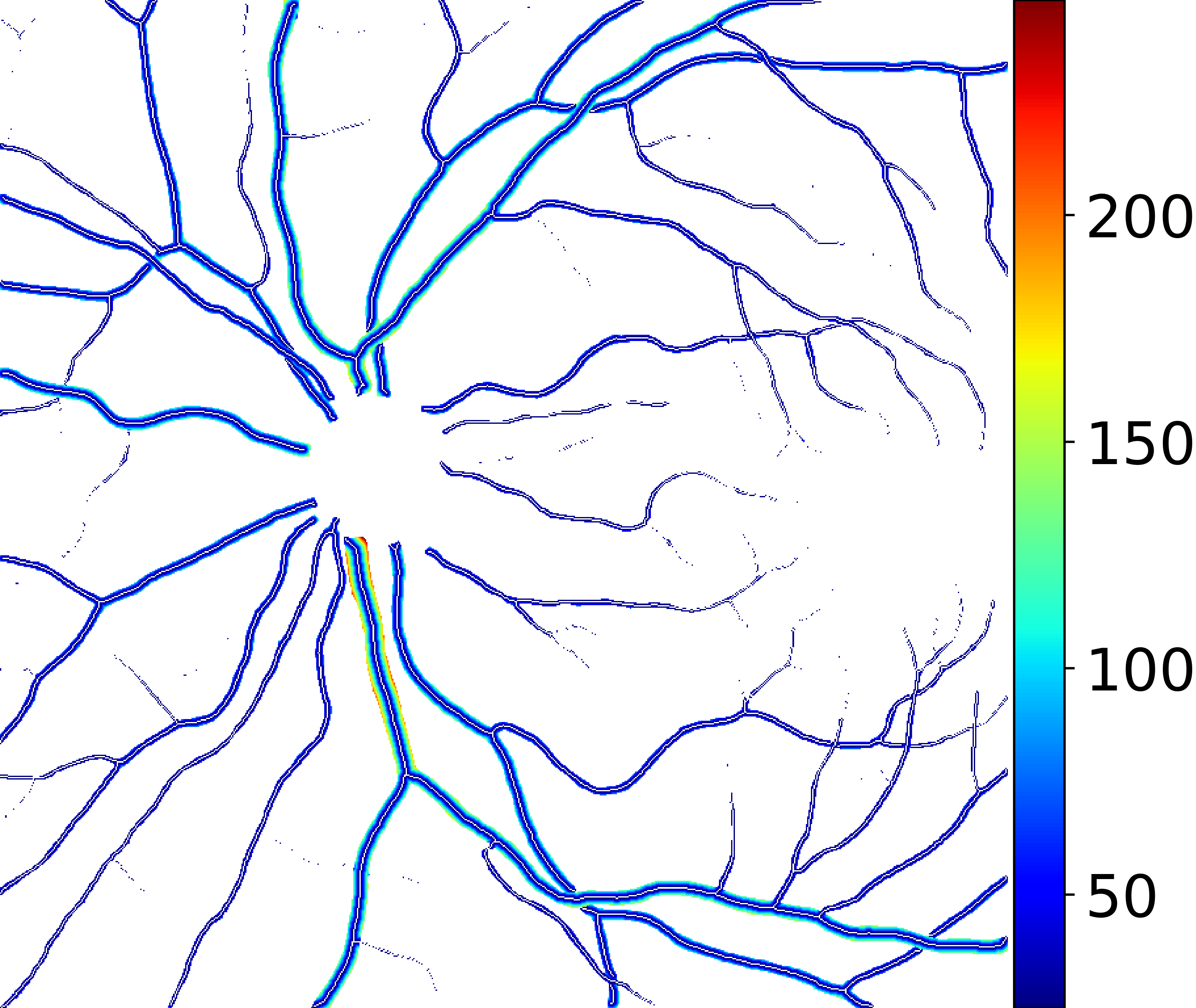}}\\[6pt]
\makebox[0.32\linewidth]{(a)}
\makebox[0.32\linewidth]{(b)} 
\makebox[0.32\linewidth]{(c)}
\caption{Qualitative comparisons of pixel-wise width maps (measured in microns), with the red and blue masks denoting artery and vein classes, respectively. (a) SegRAVIR segmentation output. (b) Reference width maps. (c) Computed width maps.}
\label{fig:width_measurefig}
\end{figure}

Table~\ref{tab_res_width} presents a quantitative comparison of the measured diameters using the segmentation outputs of SegRAVIR and competing approaches. Using the pixel-wise annotated masks, the reference average diameter of the arteries and veins in the test set of the RAVIR dataset were measured as $97.93\pm8.62$ and $113.98\pm13.93$, respectively. According to our analysis, SegRAVIR can accurately measure the diameter of the vessels and it achieves the smallest MAPE among the competing approaches. Specifically, in comparison to CE-Net, Iter-Net, and DU-Net, respectively, SegRAVIR is on average $13.20\%$, $22.73\%$ and $31.78\%$ more accurate in terms of MAPE for the measured diameter of arteries and $16.28\%$, $26.71\%$ and $27.48\%$ in terms of MAPE for the measured diameter of veins. Fig.~\ref{fig:width_measurefig} presents qualitative comparisons of reference and SegRAVIR estimated diameter maps.

\begin{figure*} \centering
\def\x{0.16}
\includegraphics[width=\x\linewidth,height=\x\linewidth]{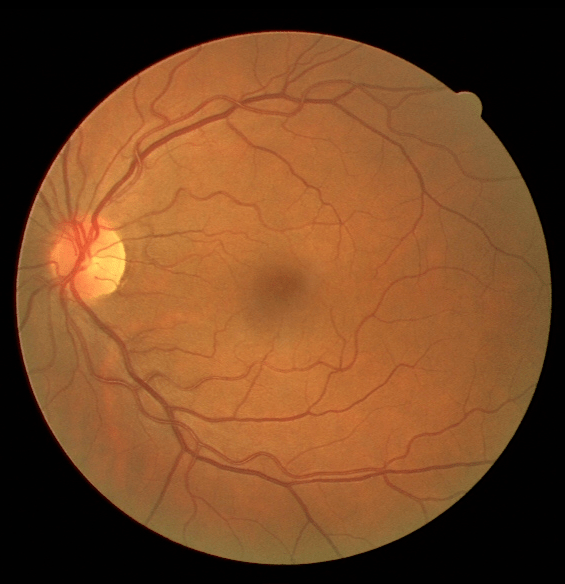}
\hfill
\includegraphics[width=\x\linewidth,height=\x\linewidth]{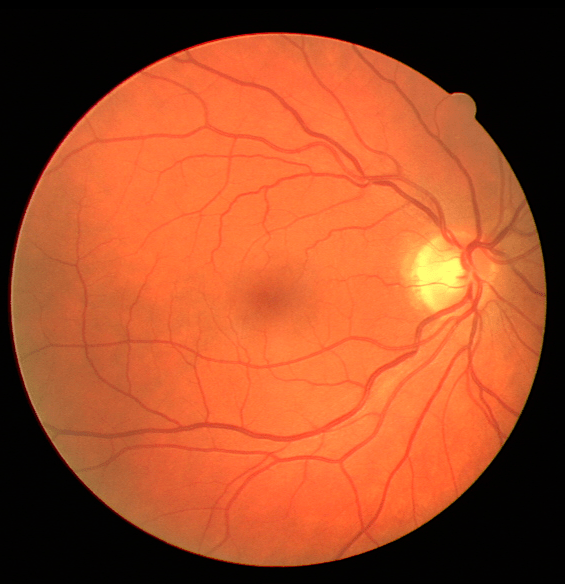}
\hfill
\includegraphics[width=\x\linewidth,height=\x\linewidth]{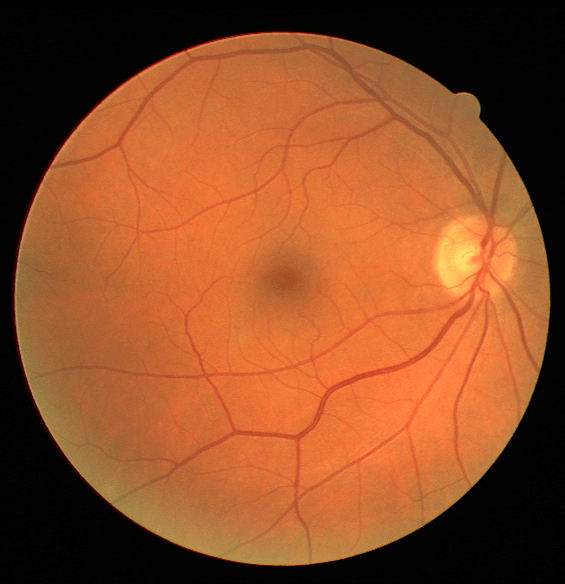}
\hfill
\includegraphics[width=\x\linewidth,height=\x\linewidth]{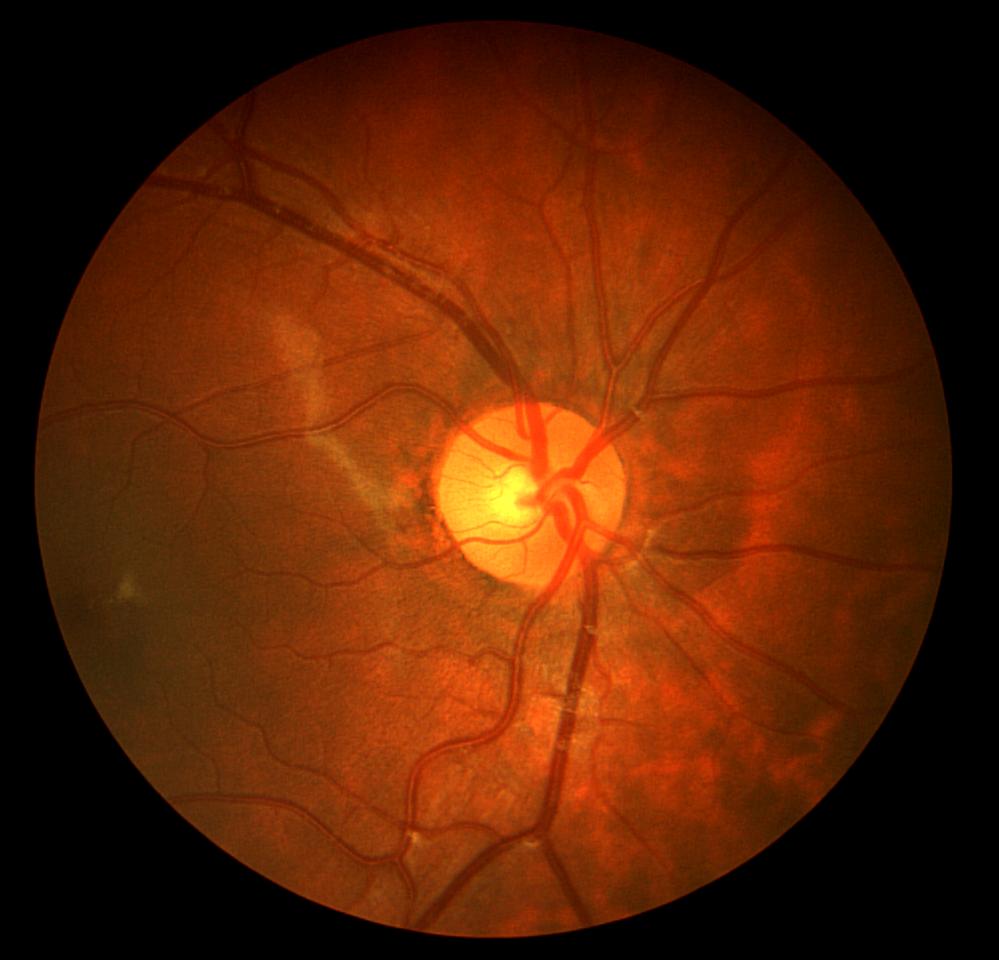}
\hfill
\includegraphics[width=\x\linewidth,height=\x\linewidth]{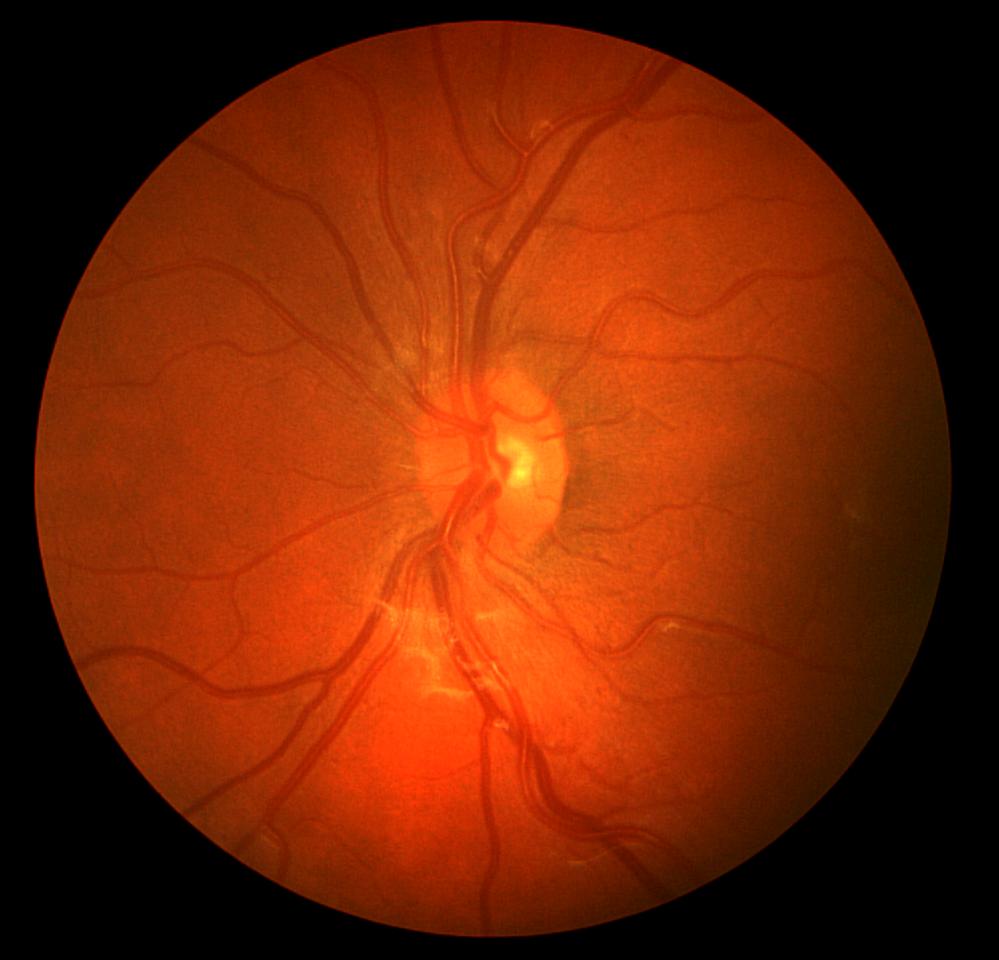}
\hfill
\includegraphics[width=\x\linewidth,height=\x\linewidth]{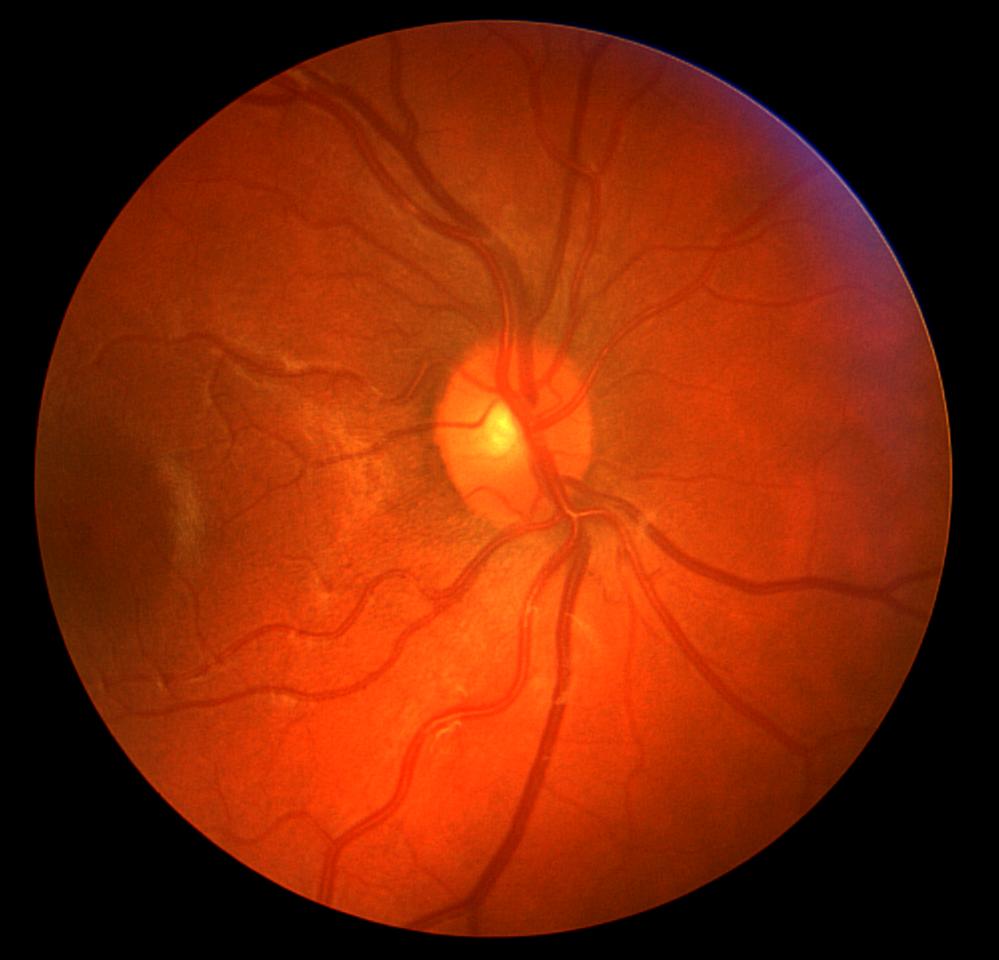}
\\[10pt]
\includegraphics[width=\x\linewidth,height=\x\linewidth]{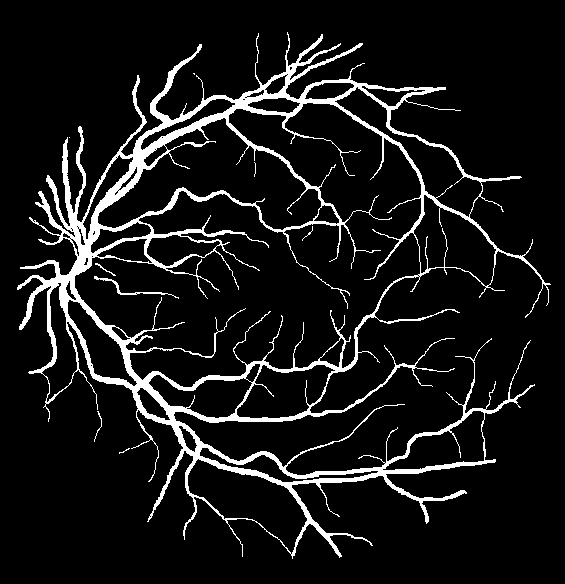}
\hfill
\includegraphics[width=\x\linewidth,height=\x\linewidth]{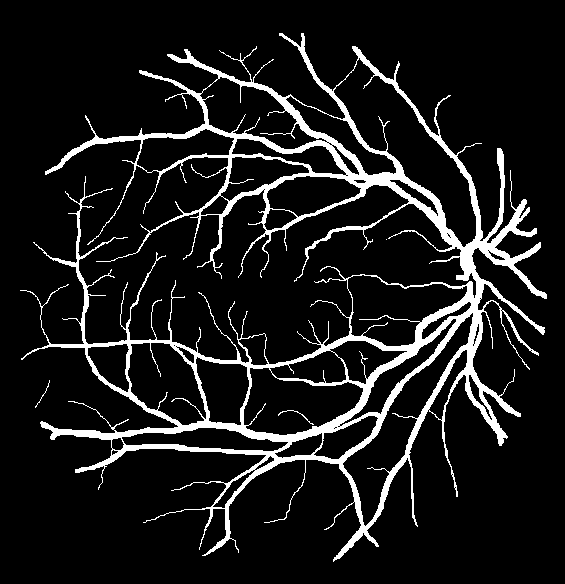}
\hfill
\includegraphics[width=\x\linewidth,height=\x\linewidth]{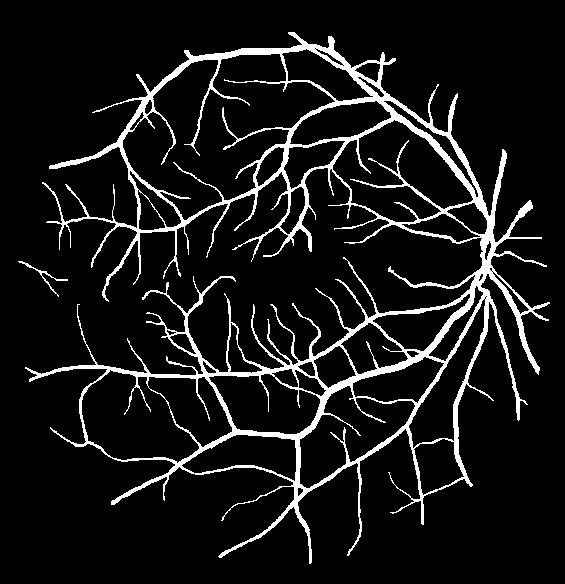}
\hfill
\includegraphics[width=\x\linewidth,height=\x\linewidth]{colored_imgs/22}
\hfill
\includegraphics[width=\x\linewidth,height=\x\linewidth]{colored_imgs/27}
\hfill
\includegraphics[width=\x\linewidth,height=\x\linewidth]{colored_imgs/28}
\\[10pt]
\includegraphics[width=\x\linewidth,height=\x\linewidth]{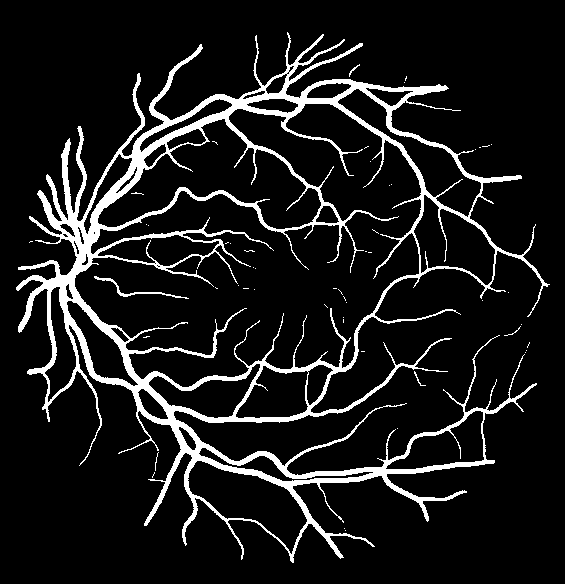}
\hfill
\includegraphics[width=\x\linewidth,height=\x\linewidth]{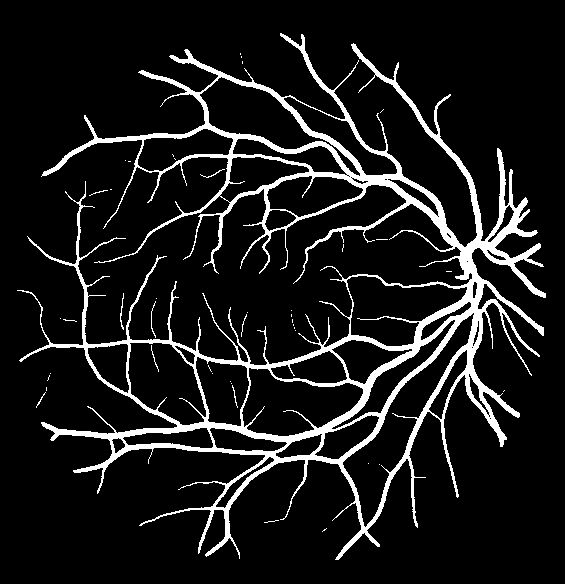}
\hfill
\includegraphics[width=\x\linewidth,height=\x\linewidth]{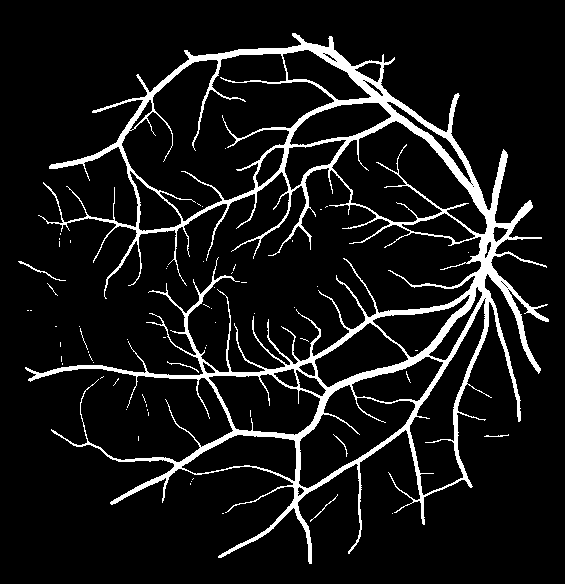}
\hfill
\includegraphics[width=\x\linewidth,height=\x\linewidth]{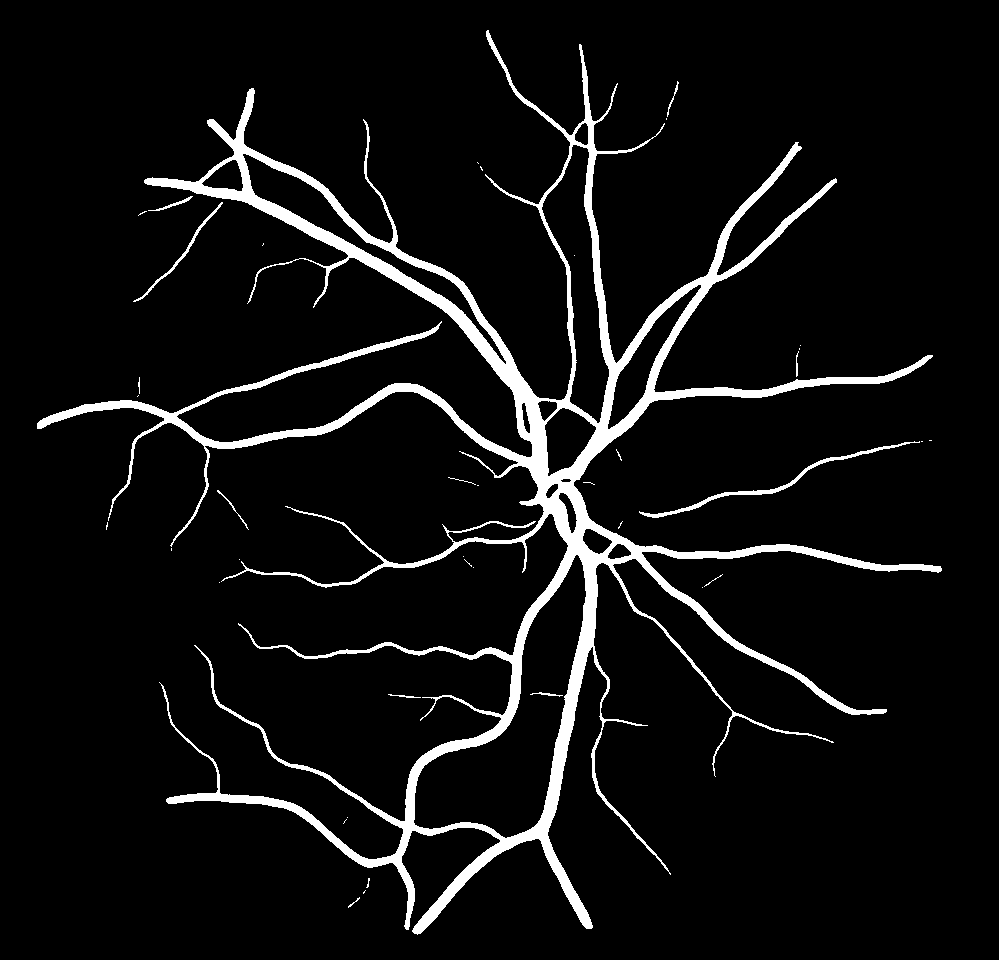}
\hfill
\includegraphics[width=\x\linewidth,height=\x\linewidth]{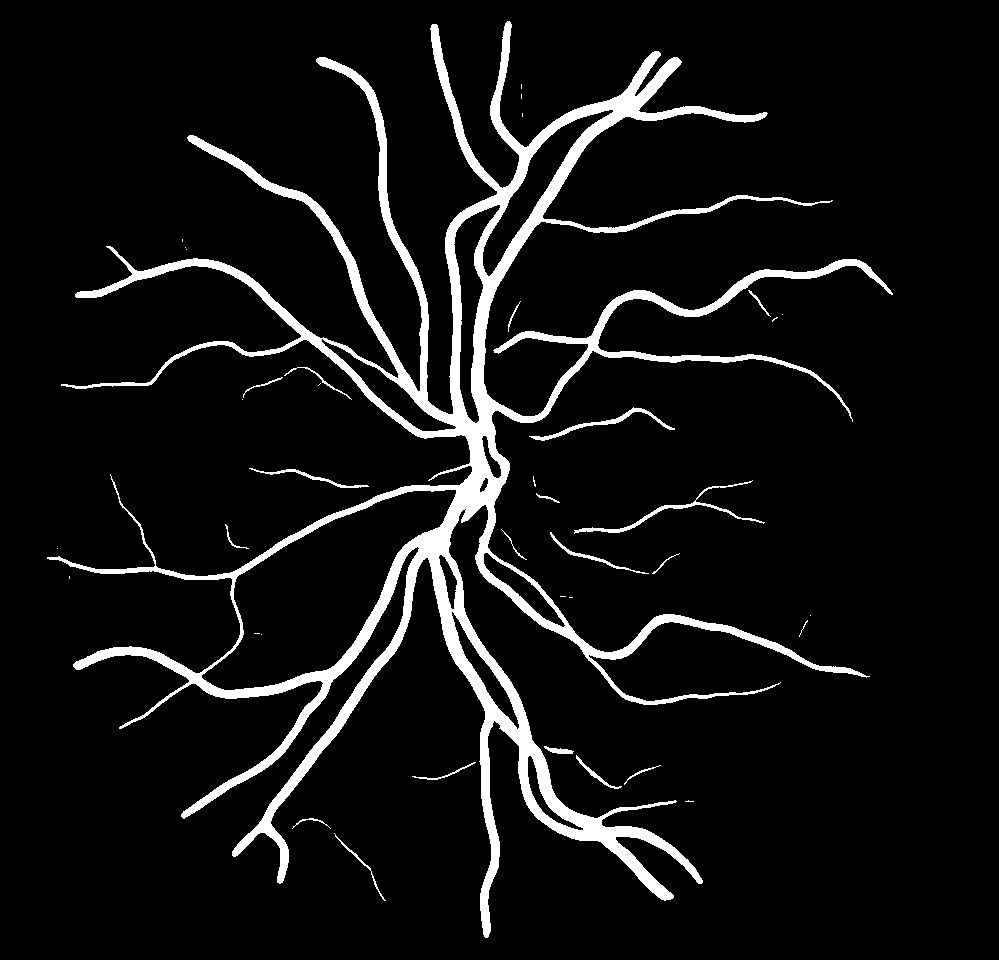}
\hfill
\includegraphics[width=\x\linewidth,height=\x\linewidth]{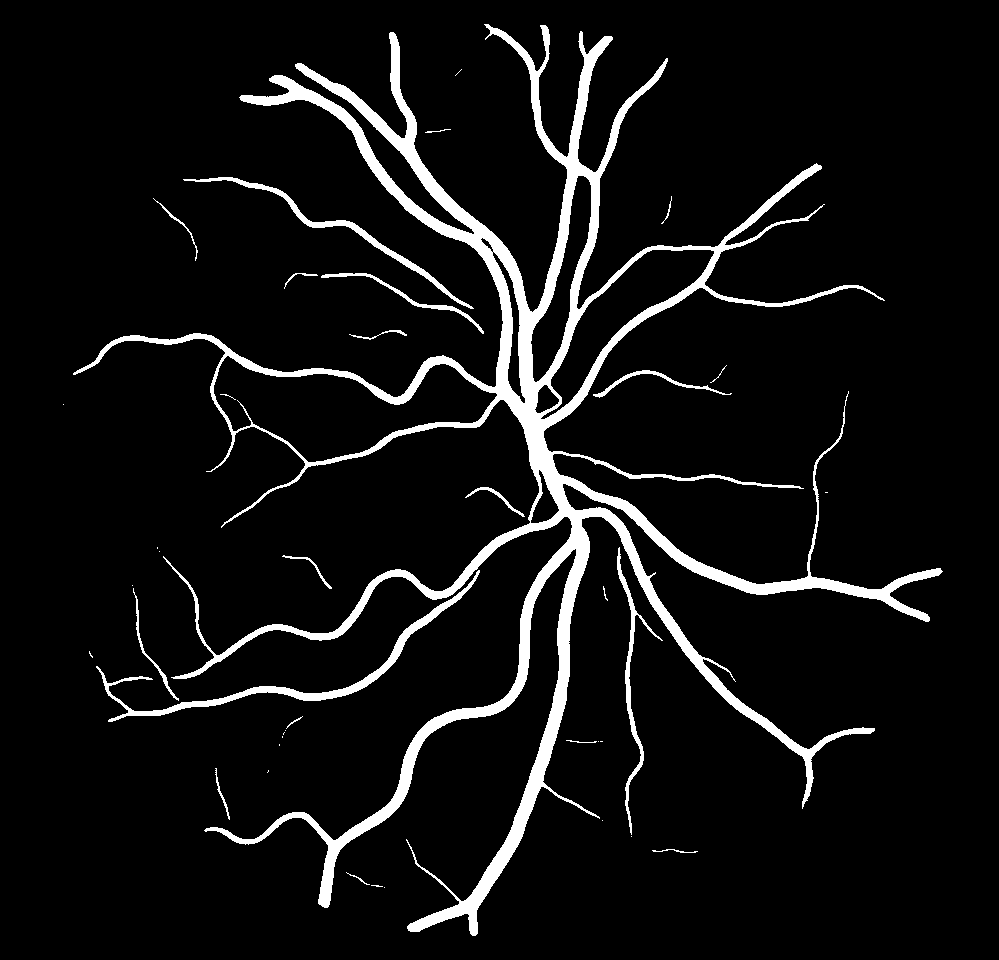}
\caption{Example SegRAVIR segmentation outputs for images in the test sets of the DRIVE (first three columns) and CHASE\_DB1 datasets. From top, the rows show input images, groundtruth labels, and segmentations.}
\label{fig:coloredseg}
\end{figure*}

\begin{table*} \centering
\caption{Quantitative comparison of segmentation performance on the DRIVE, CHASE\_DB1, and STARE datasets.}
\label{tab_res}
	\resizebox{\textwidth}{!}{
\begin{tabular}{l ccccc ccccc ccccc}
\toprule
 & \multicolumn{5}{c}{DRIVE} & \multicolumn{5}{c}{CHASE\_DB1} & \multicolumn{5}{c}{STARE}\\
\cmidrule(lr){2-6} \cmidrule(lr){7-11}  \cmidrule(lr){12-16}  
Method & SE & SP & Acc & AUC&Dice & SE & SP & Acc& AUC&Dice & SE & SP & Acc& AUC&Dice\\
\midrule
U-Net~\cite{ronneberger2015u} &  0.7822 & 0.9808 & 0.9555 & 0.9752 & 0.8174 & 0.7840 & 0.9880 & 0.9752& 0.9870&0.7993&0.6681&0.9915&0.9639&0.9710&0.7595\\
Dense U-Net~\cite{guan2019fully} &  0.7928 &0.9776& 0.9541 &0.9756 & 0.8146 & 0.7877 &0.9848& 0.9743& 0.9880&0.8005&0.6807&0.9916&0.9651&0.9755&0.7691\\
Residual U-Net~\cite{alom2019recurrent} &  0.7726 &0.9820& 0.9553 &0.9779 & 0.8149 & 0.7726 &0.9820& 0.9553& 0.9779&0.7800&0.8203&0.9856&0.9700&0.9904&0.8388\\
Recurrent U-Net~\cite{alom2019recurrent} &  0.7751 &0.9816& 0.9556 &0.9782 & 0.8155 & 0.7459 &0.9836& 0.9622& 0.9803&0.7810&0.8108&0.9871&0.9706&0.9909&0.8396\\
R2U-Net~\cite{alom2019recurrent} & 0.7792 &0.9813& 0.9556 &0.9784 & 0.8171 & 0.7756 &0.9820& 0.9634& 0.9815&0.7928&0.8298&0.9862&0.9712&0.9914&0.8475\\
DU-Net \cite{jin2019dunet} &  0.7863 &0.9805& 0.9558 & 0.9778 & 0.8190 &0.7858 & 0.9880&  0.9752& 0.9887&0.8000&0.6810&0.9903&0.9639&0.9758&0.7629\\
IterNet \cite{Li_2020_WACV} &   0.7735 & 0.9838&  0.9573 &0.9816 & 0.8205 &0.7969 & 0.9881&  0.9760& 0.9899&0.8072&0.7715&0.9886&0.9701&0.9881&0.8146\\
DAVS-Net \cite{raza2021davs} &   \textbf{0.8286} & 0.9824&  0.9689 &0.9825 & - &0.8144&0.9843&0.9726&0.9855&-&0.8238&0.9866&0.9744&0.9877&-\\
\textbf{SegRAVIR} &  0.8245 & \textbf{0.9903} &  \textbf{0.9698} & \textbf{0.9889} &  \textbf{0.8324} & \textbf{0.8172} & \textbf{0.9924} &  \textbf{0.9788} & \textbf{0.9935} &  \textbf{0.8535}&  \textbf{0.8314}&  \textbf{0.9942}&  \textbf{0.9751}&  \textbf{0.9921}&  \textbf{0.8489}\\
\bottomrule
\end{tabular}
}
\end{table*}

\begin{table} \centering
\caption{Effect of various loss functions on the segmentation performance of SegRAVIR in terms of Dice score}
\label{tab:ablation}
\begin{tabular}{llcc}
\toprule
Loss function  & Artery & Vein & Average\\
\midrule
Dice& 0.8033&0.8095&	0.8064	 \\
CE& 0.8078&0.8138&	0.8108	  \\
CE+Dice& 0.8193&	0.8225&	0.8209	  \\
CE+Dice+L2& \textbf{0.8287}&	\textbf{0.8301}&		\textbf{0.8294}	  \\
\bottomrule
\end{tabular}
\end{table}

\subsection{Knowledge Distillation}

Using the final pretrained student SegRAVIR model in our knowledge distillation framework described in Section~\ref{sec:knowledge}, we fine-tuned and tested the network on three publicly available datasets of color images, DRIVE~\cite{staal2004ridge}, STARE~\cite{hoover2000locating}, and CHASE\_DB1~\cite{fraz2012ensemble}. Fig.~\ref{fig:coloredseg} shows example of SegRAVIR segmentation outputs on DRIVE and CHASE\_DB1 datasets. Table~\ref{tab_res} presents a quantitative comparison between the segmentation performance of our SegRAVIR model and state-of-the-art models: R2U-Net~\cite{alom2019recurrent}, DU-Net~\cite{jin2019dunet}, and IterNet~\cite{Li_2020_WACV}. By all evaluation metrics, SegRAVIR has achieved new state-of-the-art results on the DRIVE, STARE~\cite{hoover2000locating}, and CHASE\_DB1 datasets. 

\section{Ablation Study}
\label{sec:abl1}

In Table~\ref{tab:ablation}, we evaluate the effectiveness of each term of SegRAVIR's loss function (\ref{eq:losstot}). Our proposed hybrid loss function, which incorporates Dice and CE losses for the main stream and L2 loss for the auxiliary stream, achieves consistently better segmentation performance for both the artery and vein classes. On average, it outperforms SegRAVIR networks trained with Dice, CE, and combined Dice and CE losses by $2.77\%$, $2.24\%$, and $1.02\%$, respectively. We hypothesize that the dominance of Dice and CE losses is due to combined improvements in artery and vein pixel-wise classification and mask prediction by the CE and Dice losses, respectively.

\begin{table} \centering
\caption{Quantitative comparison of different pretraining strategies on the DRIVE, CHASE\_DB1 and STARE datasets.}
\label{tab:ablation_v2}
\begin{tabular}{l cc cc cc}
\toprule
 & \multicolumn{2}{c}{DRIVE} & \multicolumn{2}{c}{CHASE\_DB1}&
 \multicolumn{2}{c}{STARE}\\
\cmidrule(lr){2-3} \cmidrule(lr){4-5} \cmidrule(lr){6-7} 
Method & AUC & F1 & AUC & F1& AUC & F1\\
\midrule
Random & 0.9773&0.8151&0.9817&0.7998&0.9879&0.8119\\
Pretrained & 0.9822&0.8216&0.9903&0.8297&0.9902&0.8275\\
\textbf{Proposed} & \textbf{0.9889}&\textbf{0.8324}&\textbf{0.9935}&\textbf{0.8535}&\textbf{0.9921}&\textbf{0.8489}\\
\bottomrule\\[-2pt]
\end{tabular}
\end{table}

In Table~\ref{tab:ablation_v2}, we present the segmentation performance of SegRAVIR, with different pretraining strategies, on the DRIVE, STARE~\cite{hoover2000locating}, and CHASE\_DB1 datasets. To validate the effectiveness of our proposed knowledge distillation approach, we compared against the RAVIR network pretrained on the RAVIR dataset as well as the network with random weight initialization. On the DRIVE and STARE datasets, our approach yields $2.12\%$ and $4.55\%$ improvements in Dice score, respectively, relative to random weight initialization and $1.31\%$ and $2.58\%$ relative to pretrained weight initialization. Similarly, on the CHASE\_DB1 dataset, our approach outperforms random and pretrained weight initialization by 6.71\% and 2.86\% in terms of the Dice score, respectively.

\section{Discussion}

We have tackled the problem of semantic segmentation of retinal arteries and veins in IR imagery. According to our extensive experiments, the SegRAVIR model outperforms competing approaches, evidently excelling in accurately delineating both arteries and veins. Since the latter usually manifest as larger vessels, all methods showed a better performance in segmenting veins compared to arteries. Our study reveals that the two important issues in segmenting retinal IR images are maintaining structural consistency across the vessels and correctly distinguishing between arteries and veins. SegRAVIR addresses these challenges by introducing an additional auxiliary stream that learns to reconstruct the input image via an encoder shared with the main stream and by enforcing a combined Dice and CE loss function that improves both pixel-level accuracy and contour delineation. Although, some of the competing models leverage attention modules to further localize desired retinal structures, our experiments did not reveal significant performance gains, if any.

Our experiments show that SegRAVIR can successfully segment major arteries and veins while properly capturing the fine-grained structural details of smaller vessels. Furthermore, we have quantitatively studied the problem of vessel diameter estimation using the segmentation results of our model. Results show that the segmentation outputs of SegRAVIR achieve the smallest MAPE error for both arterioles and venules and can be leveraged for precise vascular diameter analysis. In particular, such a measurement tool enables the study of morphological changes of retinal arterioles and venules and opens the door to creating early predictive models for the diagnosis of various anomalies such as diabetic and hypertensive retinopathy.

We have explored the segmentation of retinal vessels in other imaging modalities. SegRAVIR outperforms the previous state-of-the-art approaches to retinal artery and vein segmentation based on the RITE dataset~\cite{hu2013automated}, hence validating the generalizability of our approach to other datasets. In addition, we proposed a knowledge distillation approach to domain adaptation of RAVIR-pretrained models. Although naive pretraining on the RAVIR dataset improves the segmentation performance on color datasets, our approach significantly improves this performance baseline. As a result, our model establishes new state-of-the-art performance benchmarks on the DRIVE\cite{staal2004ridge}, STARE~\cite{hoover2000locating}, and CHASE\_DB1~\cite{fraz2012ensemble} datasets. This indicates that our RAVIR dataset can be leveraged to address the annotated data scarcity problem in this domain, and our SegRAVIR model can be effectively used for multi-modal segmentation across different domains.

\section{Conclusions}

In summary, RAVIR is a new retinal IR image dataset that has engendered SegRAVIR, a novel, deep learning model for the fully-automated semantic segmentation and quantitative analysis of retinal vasculature. SegRAVIR can accurately segment both arteries and veins without additional preprocessing or post-processing, and its output enables precise vessel diameter mensuration critical to the quantitative assessment of retinal vasculature. Furthermore, via a knowledge distillation approach, our dataset and model can be effectively applied to retinal vasculature segmentation based on other imaging modalities. This enables new opportunities in retinal image analysis while helping alleviate the problem of annotated data scarcity in this domain. 

\balance

\bibliographystyle{IEEEtran}
\bibliography{ref}

\end{document}